\documentclass[sigconf,authorversion]{acmart}

\usepackage{multirow}

\AtBeginDocument{%
  \providecommand\BibTeX{{%
    \normalfont B\kern-0.5em{\scshape i\kern-0.25em b}\kern-0.8em\TeX}}}

\copyrightyear{2025}
\acmYear{2025}
\setcopyright{cc}
\setcctype{by}
\acmConference[CHI '25]{CHI Conference on Human Factors in Computing Systems}{April 26-May 1, 2025}{Yokohama, Japan}
\acmBooktitle{CHI Conference on Human Factors in Computing Systems (CHI '25), April 26-May 1, 2025, Yokohama, Japan}\acmDOI{10.1145/3706598.3713295}
\acmISBN{979-8-4007-1394-1/2025/04}

\begin{document}

\title[AI, Help Me Think---but for Myself]{AI, Help Me Think---but for Myself: Assisting People in Complex Decision-Making by Providing Different Kinds of Cognitive Support}

\author{Leon Reicherts}
\authornote{Both authors contributed equally to this research.}
\email{leon.reicherts@microsoft.com}
\orcid{0000-0002-7338-5242}
\affiliation{%
 \institution{Microsoft Research}
 \city{Cambridge}
 \country{United Kingdom}
}
\affiliation{%
 \institution{University College London}
 \city{London}
 \country{United Kingdom}
}

\author{Zelun Tony Zhang}
\authornotemark[1]
\email{zhang@fortiss.org}
\orcid{0000-0002-4544-7389}
\affiliation{%
  \institution{fortiss GmbH, Research Institute of the Free State of Bavaria}
  \city{Munich}
  \country{Germany}
}
\affiliation{%
  \institution{LMU Munich}
  \city{Munich}
  \country{Germany}
}

\author{Elisabeth von Oswald}
\email{elisabethluisa.von@mail.polimi.it}
\orcid{0009-0004-7840-351X}
\affiliation{%
  \institution{Politecnico di Milano}
  \city{Milan}
  \country{Italy}
}

\author{Yuanting Liu}
\email{liu@fortiss.org}
\orcid{0000-0002-8651-6272}
\affiliation{%
  \institution{fortiss GmbH, Research Institute of the Free State of Bavaria}
  \city{Munich}
  \country{Germany}
}

\author{Yvonne Rogers}
\email{y.rogers@ucl.ac.uk}
\orcid{0000-0001-7612-0234}
\affiliation{%
  \institution{University College London}
  \city{London}
  \country{United Kingdom}
}

\author{Mariam Hassib}
\email{hassib@fortiss.org}
\orcid{0000-0001-6530-9357}
\affiliation{%
  \institution{fortiss GmbH, Research Institute of the Free State of Bavaria}
  \city{Munich}
  \country{Germany}
}

\renewcommand{\shortauthors}{Reicherts and Zhang, et al.}

\begin{abstract}

    How can we design AI tools that effectively support human decision-making by complementing and enhancing users' reasoning processes? Common recommendation-centric approaches face challenges such as inappropriate reliance or a lack of integration with users' decision-making processes. Here, we explore an alternative interaction model in which the AI outputs build upon users' own decision-making rationales. We compare this approach, which we call ExtendAI, with a recommendation-based AI. Participants in our mixed-methods user study interacted with both AIs as part of an investment decision-making task. We found that the AIs had different impacts, with ExtendAI integrating better into the decision-making process and people's own thinking and leading to slightly better outcomes. RecommendAI was able to provide more novel insights while requiring less cognitive effort. We discuss the implications of these and other findings along with three tensions of AI-assisted decision-making which our study revealed.
\end{abstract}



\begin{teaserfigure}
    \centering
  \includegraphics[width=0.93\textwidth]{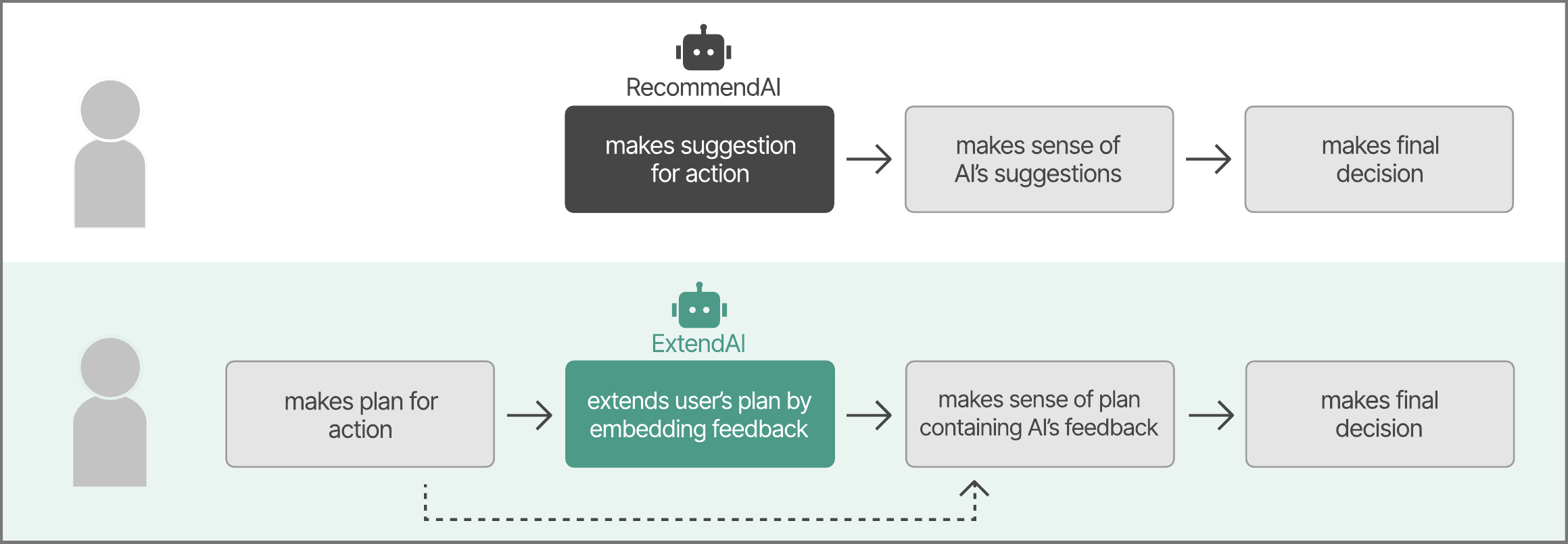}
  \caption{Illustrative comparison of the thought process when interacting with two `types' of AI -- \textit{RecommendAI} and \textit{ExtendAI}.}
  \Description{A comparison of the user interaction and decision-making processes with the RecommendAI and ExtendAI. The flowchart is divided into two rows, each representing one of the AI assistance types. For RecommendAI the AI "makes a suggestion for action". The user then "makes sense of the AI's suggestions" and moves on to "making a final decision". For ExtendAI, the user "makes a plan for action" after which the AI "extends the user's plan by embedding feedback" into it. The user then proceeds to "making sense of the plan containing Al's feedback" to then "make the final decision".}  \hspace{1em}
  \label{fig:teaser}
\end{teaserfigure}

\maketitle

\section{Introduction}
AI is increasingly being used to support human decision-making in a variety of domains, such as healthcare~\cite{secinaroRoleArtificialIntelligence2021}, finance~\cite{caoAIFinanceChallenges2023}, and logistics~\cite{woschankReviewFurtherDirections2020}. Typically, this involves an interaction model where the AI makes recommendations, which the user can then take into account in their decision-making. Benefits of \textit{AI-assisted decision-making} include both increased decision efficiency and increased accuracy~\cite{baoLiteratureReviewHuman2023}. However, various challenges persist, including inappropriate reliance on AI~\cite{bansalDoesWholeExceed2021,bucincaTrustThinkCognitive2021,greenPrinciplesLimitsAlgorithmloop2019,poursabzi-sangdehManipulatingMeasuringModel2021,dietvorstAlgorithmAversionPeople2015,chengOvercomingAlgorithmAversion2023}, deskilling~\cite{suttonHowMuchAutomation2018,gajosPeopleEngageCognitively2022}, or the difficulty of integrating AI recommendations into users' decision-making process~\cite{blombergActingAnalyticsAccuracy2018,kawakamiWhyCareWhat2022,zhangRethinkingHumanAICollaboration2024}. This difficulty often arises when factors outside of the features modelled by an AI system influence the decision, which \citet{holsteinSupportingPerceptualComplementarity2023} call ``unobservables''. This is particularly challenging when decisions involve unstructured human reasoning and context, such as complex family backgrounds in a social care setting~\cite{kawakamiWhyCareWhat2022}, or patient preferences in healthcare~\cite{jacobsDesigningAITrust2021}, which AI systems traditionally have not been able to capture.

Most work in AI-assisted decision-making attempts to address these challenges while remaining within the common paradigm of providing recommendations. 
For instance, a large and rapidly growing body of work revolves around supplementing recommendations with explanations of how the AI produced its recommendation~\cite{laiScienceHumanAIDecision2023}. However, it has been argued in recent work that at least some of the challenges are inherent to the recommendation-centric paradigm, and proposed to support human decision-making by other means than recommendations~\cite{zhangRethinkingHumanAICollaboration2024,zhangRecommendationsBackwardForward2024,millerExplainableAIDead2023}. The proposition is to augment users' reasoning while they reason \textit{forward} through the decision-making task toward their own solution, rather than recommending a solution from which users have to reason \textit{backward}~\cite{zhangRecommendationsBackwardForward2024,wangDesigningTheorydrivenUsercentric2019}.

To this end, with their unique capabilities, large language models (LLMs) could enable various yet unexplored roles for AI to augment human decision-making~\cite{eignerDeterminantsLLMassistedDecisionmaking2024}. For one, LLMs have been found to be highly proficient in various expert domains such as medicine~\cite{noriCanGeneralistFoundation2023} and finance~\cite{fiebergUsingGPT4Financial2023,kimFinancialStatementAnalysis2024}, often surpassing the performance of human domain experts. More importantly, the impressive natural language capabilities of LLMs may provide new opportunities to flexibly process unstructured data and human reasoning. This provides new opportunities to more tightly integrate AI into users' decision-making process, especially in complex, ill-defined problem settings. Recent research explored utilising these capabilities of LLMs to help users engage more deeply with AI recommendations~\cite{maHumanAIDeliberationDesign2024,chiangEnhancingAIassistedGroup2024,yangHarnessingBiomedicalLiterature2023}. Our research is concerned with how we can leverage the natural language capabilities of LLMs to provide a new tool to support human reasoning. In particular, we investigated how an LLM can be integrated with a software tool in order to elicit forward reasoning for a complex task; in this case, we chose financial investments, where many decisions need to be made. 
Our aim is to embed more fine-grained AI support into users' own reasoning, allowing users to remain engaged with the decision-making task, rather than diverting their engagement to AI recommendations. We do so by asking users to write a rationale for their decision and extending it with LLM-generated feedback, meant to help users reflect on, and identify potential gaps in their reasoning (see \autoref{fig:teaser}). Given that this represents a different paradigm than the usual recommendation-focused approach, we pose the following research question:

\begin{quotation}
    \textbf{RQ:} How do different paradigms for AI decision support---one that extends users' reasoning and one that provides direct recommendations---affect users’ decision-making processes, perceptions of the AI, and decision outcomes in complex decisions?
\end{quotation}

 Our chosen use case, financial investment, presents a complex and realistic test bed for our exploration. In a mixed methods study ($N$=21), we find that both approaches can support human decision-making in different ways. Our findings show that the approach of extending people's rationales helps them reflect on their reasoning, leading to more informed and better decisions. On the other hand, the recommendation-based approach seems to help people get inspiration for and explore new directions in their decision-making. We found marked differences and tensions in the preferences for both approaches depending on the participant's personality, expertise, and decision-making style, resulting in an even split of overall preference between the two designs.

The contributions of this work are threefold: (1) the conceptualisation and design of two LLM roles for AI decision support in the context of a realistic, complex, and open-ended task, (2) an exploration of the perceptions and usage of the LLM assistance through a mixed-methods study within a complex financial investment use case, (3) a discussion of the tensions of integrating AI into complex decision-making tasks.

\section{Background \& Related Work}
In this section, we first review the broader literature on AI-assisted decision support followed by an overview of research that specifically uses LLMs for this purpose.

\subsection{AI-Assisted Decision-Making}
\label{sec:AI-assisted-DM}
The ultimate goal of AI decision support tools is to work \emph{with} humans to potentially reach the best performance that overcomes the limitations of either human or AI on their own. This is often referred to as human-AI complementarity \cite{bansalDoesWholeExceed2021}. Decision-making is a highly complex process and AI support tools may vary profoundly in when, how, and what they provide. Research in AI-assisted decision-making largely focuses on providing AI-generated recommendations from model predictions (see \citet{laiScienceHumanAIDecision2023, baoLiteratureReviewHuman2023} for reviews).  Users must then evaluate and consequently accept or reject the recommendation~\cite{laiScienceHumanAIDecision2023}. Recommendations are either presented on their own or combined with additional AI support elements such as various types of model explanations~\cite{linardatosExplainableAIReview2020,guidottiSurveyMethodsExplaining2018} or uncertainty~\cite{zhangEffectConfidenceExplanation2020,prabhudesaiUnderstandingUncertaintyHow2023,bansalDoesWholeExceed2021}. While sometimes successful in increasing accuracy and efficiency in generic or artificial tasks~\cite{yangHowVisualExplanations2020,laiDeception,Vasconcelos2023}, AI-assisted decision-making has its challenges, including low user acceptance in more complex decision-making processes~\cite{kawakamiWhyCareWhat2022,blombergActingAnalyticsAccuracy2018,zhangRethinkingHumanAICollaboration2024}, and overreliance on (incorrect) AI recommendations~\cite{bansalDoesWholeExceed2021, Bussone2015,burgessHealthcareAITreatment2023,jacobsHowMachinelearningRecommendations2021}. %
Empirical results on the effectiveness of explanations or other support elements in calibrating reliance have been mixed. While there are instances where explanations lead to more appropriate reliance~\cite{Vasconcelos2023,yangHowVisualExplanations2020,wangHelpful}, in many cases, explanations have had little effect on users' decision accuracy \cite{cabitza2024never} or even increased overreliance~\cite{bansalDoesWholeExceed2021, Bussone2015,eiband2019, jacobsHowMachinelearningRecommendations2021,schmidtCalibratingHumanAICollaboration2020,bertrandQuestioningAbilityFeaturebased2023,schemmerAppropriateRelianceAI2023, cabitza2024explanations}. %

Another body of emerging research investigates more complex, realistic decision-making tasks, which require humans to work with subjective and unstructured knowledge going beyond single outcome predictions, such as in healthcare, social welfare, or financial decision-making~\cite{bucincaEndPredictionsStop2022,koonHumancapabilitiesOrientationHumanAI2022,zhangForwardReasoningDecision2021}. Real-world decision tasks often call for a more nuanced type of AI decision support that accompanies the user \textit{during} the different phases of their decision-making and \textit{augments} the process \cite{zhangForwardReasoningDecision2021,zhangRecommendationsBackwardForward2024}, instead of providing one concrete prediction as \textit{the} ground truth solution. One example is the concept of \textit{Evaluative AI} proposed by \citet{millerExplainableAIDead2023}, where AI helps users generate and evaluate hypotheses, while leaving the decision itself to users. \citet{zhangRethinkingHumanAICollaboration2024} applied a similar concept in their redesign of an existing AI-based sepsis diagnosis tool, aiming to support the process leading up to a decision. 
Another example is \textit{Reflective XAI} by \citet{cabitza2023let}, which prompts radiologists to reflect on the decisions of the AI by presenting evidence for and against the AI advice.
In a study on an aviation use case~\cite{zhangRecommendationsBackwardForward2024}, the main challenge for pilots was integrating data from multiple sources rather than interpreting that data. AI thus proved valuable by presenting relevant information at a glance and intelligently highlighting constraints at nearby airports, instead of recommending a concrete airport. 
\citet{yildirimMultimodalHealthcareAI2024} similarly found that radiologists preferred AI to help them complete tedious sub-tasks or collect relevant evidence for their decisions rather than taking over the interpretation of the evidence. Even more restrained approaches where the AI completely refrains from analysis tasks have also been explored, for instance, through chatbots that support decision-making through context-dependent probing questions to scaffold users' reasoning~\cite{reichertsExtendingChatbotsProbe2022}. 

Overall, decision-making tasks are dispersed along a space of complexity that requires a flexible approach to AI support. To understand which designs of AI decision support are suited for which decision task, it is crucial to empirically compare these different designs. One such study is that of \citet{maRecommenderExploratoryStudy2024}, who investigated the effectiveness of direct recommendations compared to two alternative designs. In one design, the AI presented a balanced analysis by providing the features speaking for and against each possible answer, aligning with the evaluative AI framework discussed by \citet{millerExplainableAIDead2023}. In the other design, the AI challenged the user's preferred choice by highlighting the features speaking against the user's answer. Results showed that each approach has its unique advantages and disadvantages, and that recommendations are not always the most effective approach. However, the simplicity of the text classification task used in the study might conceal some nuances in real-world applications. Another study exploring non-recommendation-based approaches within the context of a  more realistic use case, aviation, is that of \citet{zhangRecommendationsBackwardForward2024}. They conducted a comparison between recommendations and the continuous support approach described above. They found that continuous support was better accepted by pilots and induced less overreliance. A combination of recommendations and continuous support led to faster decisions.

\subsection{LLMs in AI-Assisted Decision-Making}
Research on the use of LLMs for AI-assisted decision-making is still in its early stages, with the few existing studies mainly following three broad directions. The first is applying LLMs within the established paradigm of explainable AI. For instance, \citet{mavrepisXAIAllCan2024} proposed creating more approachable natural language-based explanations. \citet{mozannarEffectiveHumanAITeams2023} use natural language to describe to users when to rely on AI recommendations and when not. A second direction is the usage of LLMs to enrich AI recommendations with additional evidence that is related to the task rather than the AI model. This has been investigated in a clinical application by \citet{yangHarnessingBiomedicalLiterature2023}, who used an LLM to identify supporting and opposing evidence for AI suggestions from biomedical literature. Lastly, researchers have explored the use of LLMs for designing AI agents that can engage with users in various forms of discussions. For example, \citet{chiangEnhancingAIassistedGroup2024} employed LLMs as Devil's Advocate, where the LLM debates either against the AI recommendation or the majority opinion in a group decision-making setting. \citet{maHumanAIDeliberationDesign2024} on the other hand proposed the concept of \textit{Deliberative AI}, where both the human and the AI first present their viewpoints. Afterwards, they deliberate on conflicting opinions and discuss individual features until a final decision is reached. This flow, in which the user starts by presenting their own rationale and then receives the AI's perspective, is similar to our proposed approach. However, our approach differs from this framing in that we propose that the AI extends and embeds into the human's rationale, aiming to engage the user in a forward reasoning manner. Hence, the user does not have to make sense of an independent AI viewpoint.

Overall, while the above approaches differ significantly from each other, all of them are focused on helping users to make sense of and appropriately rely on AI support received as recommendations. Additionally, most of the studies above were based on simple or generic decision-making tasks. In our work, we aim to explore how LLMs can assume a role that accompanies the user in their own reasoning during decision-making in a realistic task and how such a role compares to the more common role of providing recommendations.

\section{Use Case and Prototype}
\label{sec:use_case}
As a use case for our study we chose financial investments, as it represents a highly complex and open-ended decision problem that allowed us to explore different ways of helping the user's thought process when making decisions. Countless potentially interacting factors can play a role in a person's investment decisions, and there is generally no objective right or wrong for how to compose a portfolio. According to modern portfolio theory~\cite{markowitzPortfolioSelection1952}, an optimal portfolio is one that maximises return at the risk level that is acceptable to the investor and is thus highly individual. The primary means to reduce risk is through diversification of the portfolio. However, people often struggle to assess the risk of an investment~\cite{bradburyImprovingInvestmentDecisions2014}. Inexperienced investors also often misjudge the diversification of their portfolio as they neglect covariations~\cite{hedesstromCovariationNeglectNovice2006,laudenbachHowAlleviateCorrelation2023}, i.e., the value of individual assets might be correlated, and the portfolio thus less diversified than thought. At the same time, initial analyses suggest that LLMs like GPT-4 can perform impressively well on financial tasks~\cite{fiebergUsingGPT4Financial2023,kimFinancialStatementAnalysis2024}. These aspects and opportunities made investment decision-making a promising scenario to explore the use of LLMs for investment decision support.

The type of financial product that our study focused on were exchange-traded funds (ETFs), which are investment funds traded on stock exchanges like individual stocks. ETFs hold a range of financial assets, including stocks, bonds, currencies, debts, or commodities such as gold or crude oil. The main reason for choosing ETFs over trading other financial products was that they require less specific expert knowledge (e.g., valuation techniques for stocks; knowledge of leverage, margin requirements, and precise market timing for futures contracts; or assessing creditworthiness, interest rates, and repayment structures for bonds) and also keep the task simpler and more manageable for a study. We further simplified the task by only including ETFs that are composed of stocks (also referred to as \textit{equity ETFs}). Participants could thus focus in their reasoning and decision-making on considerations such as the regions, sectors, or themes (e.g., energy storage) to include in their portfolio, making the task more approachable to participants, while still keeping the complexity of having to weigh off different investments against each other and ways to compose them in a portfolio. Rather than asking participants to put together a portfolio for themselves, we asked them to put themselves into the shoes of a 40-year-old, rather cautious person with a long-term investment horizon of about 10 years, with the explicit goal of creating a diversified portfolio. The purpose of this was to make it less likely that participants just bought their `go-to' ETFs without much thought.

We designed a simulated stock trading platform in which we embedded an AI investing `assistant' \textbf{offering two different types of interaction/assistance}: One follows the more `traditional' approach of \textbf{making direct recommendations}. The other requires the user to describe their own rationale first. The AI assistant would then \textbf{extend the user's rationale} by embedding its feedback into it. In what follows, we will refer to both types of assistants as \textbf{\textit{RecommendAI}} and \textbf{\textit{ExtendAI}}. In the present investing context, \emph{RecommendAI} would provide a set of ETFs tailored to the user's portfolio at the given time, and \emph{ExtendAI} would ask the user to describe a rationale for how they might adjust their portfolio (i.e., which ETFs in their portfolio they intend to increase or reduce their exposure to, sell completely, or which new ones they intend to buy and why), which the AI would then give feedback on. This `augmented' rationale would then be returned containing the AI's feedback in bold so that the user could see the AI's suggestions directly in the context of their initial thoughts/reasoning (see also \autoref{fig:teaser} for a simplified overview of the decision-making process with both AIs). 
In the following section, a more detailed description of the prototype implementation will be provided.

\section{Prototype Implementation}
This section describes the design and development of the prototype in more detail. \autoref{sec:data-pre-study} outlines how the ETF dataset was prepared, and \autoref{sec:synthesising-data} describes how additional synthetic data was generated. This is followed by an overview of the design of the simulated trading platform and the AI assistants in \autoref{sec:interface-and-ai-design}.

\subsection{Preparation of ETF Data and Pre-Study}
\label{sec:data-pre-study}
In order to investigate people's decision-making on which ETFs to put together in a portfolio, we first collected and prepared an ETF dataset and then tested the suitability of the dataset for such a decision-making task in an online pre-study ($N$=10). Both these steps are described in this section.

\subsubsection{Collection and Preparation of ETF Data}
\label{sec:etf-data}
We selected a range of 31 ETFs using the ETF Screener from ETF Database\footnote{\url{https://etfdb.com/screener/}} to present in our simulated trading platform. We aimed to have enough ETFs to make the decision complex, but not too many that it overwhelms or gets too time-consuming for participants to go through them. We selected ETFs reflecting all major regional markets, all sectors from the Global Industry Classification Standard (GICS)\footnote{\url{https://en.wikipedia.org/wiki/Global_Industry_Classification_Standard}}, and popular investment strategy-related themes (e.g., low volatility, wide moat, momentum, quality, small size, value), such that common categories of ETFs would be well covered. For each category, we aimed to select one ETF, to limit decisions to choosing among these categories, not among ETFs within the same category. For sector and theme ETFs, the goal was to select global ETFs. For cases when there was no global ETF, we chose a US-focused ETF in addition to its ex-US counterpart, allowing participants to express a global view for that category. In addition, we included narrow `trend' ETFs for some sectors, such as biotechnology or clean energy ETFs for the healthcare and energy sectors, respectively. 

We used the Python package \texttt{financetoolkit}\footnote{\url{https://www.jeroenbouma.com/projects/financetoolkit}} to retrieve historical closing prices for each month of the past five years, and \texttt{thepassiveinvestor}\footnote{\url{https://www.jeroenbouma.com/projects/thepassiveinvestor}} for information such as sector and company holdings as well as performance metrics for each ETF to display in our simulated trading platform. We further retrieved country holdings for each ETF using the Financial Modeling Prep API\footnote{\url{https://site.financialmodelingprep.com/developer/docs}} and manually retrieved further details such as short analyses from the ETF Database that would be used for prompting our AI investment assistants (\autoref{sec:interface-and-ai-design}).

\subsubsection{Pre-Study to Evaluate the Decision-Making Task \& Iterate on Prompting}
\label{sec:pre-study}
After collecting and preparing our ETF dataset, we conducted a pre-study with $N$=10 participants over Prolific\footnote{\url{https://www.prolific.com/}}. Participants were provided with the ETF data and were tasked with putting together a portfolio to match the provided investor profile and writing a rationale explaining their choices. The pre-study aimed to assess: (1)~how long it would take people with some ETF investing experience to analyse the data for the 31 ETFs, compose a portfolio, and write a rationale motivating their decisions, (2)~whether the written rationales would be sufficiently detailed and elaborate, and finally (3)~how well an LLM would be able to give feedback on these rationales and how the prompts would need to be designed for best results. 
We chose pre-screeners to only select participants with some investing experience. Participants completed the task in about 8 minutes (median duration) and were able to write sufficiently elaborate rationales. We used these rationales to assess the ability of GPT-4o to provide meaningful feedback for the kind of assistance we intended to achieve with \emph{ExtendAI}.  We iterated accordingly on our metaprompt to lead to feedback that (a) considers each aspect mentioned by the users in the rationale, (b) is based on the provided investor profile, (c) does not provide specific recommendations, and (d) addresses diversification, the return and risk balance, volatility, and potential overlaps in the intended trades (between regions and sectors). This seemed to work sufficiently well, but occasionally, the LLM would not adhere to (c) and make recommendations for specific ETFs. We also added several constraints to the prompt such as ensuring that the user provided in their rationale at least two ETFs and reasons for choosing them and that only the list of the available 31 ETFs would be considered. In \autoref{app:prompts} we provide the final metaprompt we arrived at. %

\subsection{Finalisation and Synthesis of the ETF Data for the Simulated Trading Platform}
\label{sec:synthesising-data}
Following the encouraging findings of the pre-study, we further developed the study materials. We set the study in a hypothetical future rather than in the past to avoid that participants would rely on their memory of events (e.g., the stock tech boom within the first year of the pandemic or with the launch of ChatGPT, etc.) and hence not engage in considerations such as balancing risks with potential returns. This meant that synthetic data had to be generated for this hypothetical future.

When generating the synthetic ETF data, we aimed to create plausible and engaging future developments rather than the most realistic ones. The data consisted of two interdependent components:  three `news articles' describing market developments, and synthetically generated price data for each of the 31 ETFs that are in line with these news articles for August 2024, August 2026, and August 2028, the three time steps presented in the study task.

To create the `news articles', three authors had a brainstorming session on possible future economic and stock market developments while considering real articles published by financial institutions, investment platforms, and newspapers. One of the authors further expanded some of these developments and presented the drafts back to two co-authors. Based on this discussion, a final version was created, highlighting key developments such as technological breakthroughs, industrial shifts, supply chain disruptions, and rising demand for specific services, raw materials, and products---relevant to different regions, industries, and ETF themes (see supplementary materials).
We then adjusted the percentage price changes for each of the 31 ETFs to be adequately aligned with the narrative of 2026 and 2028. For an additional plausibility check, the specified percentage changes along with the news articles were reviewed by an investment banking expert with over 10 years of experience in investing in the stock market and economics. Based on the feedback confirming the sufficient plausibility of the news articles and the percentage changes, final adjustments were made. The next section describes the design and development of the simulated ETF trading platform and the assistants.

\subsection{Design of the Simulated Investing Platform and the AI Assistants}
\label{sec:interface-and-ai-design}

\begin{figure*}
    \centering
    \includegraphics[width=0.9\linewidth]{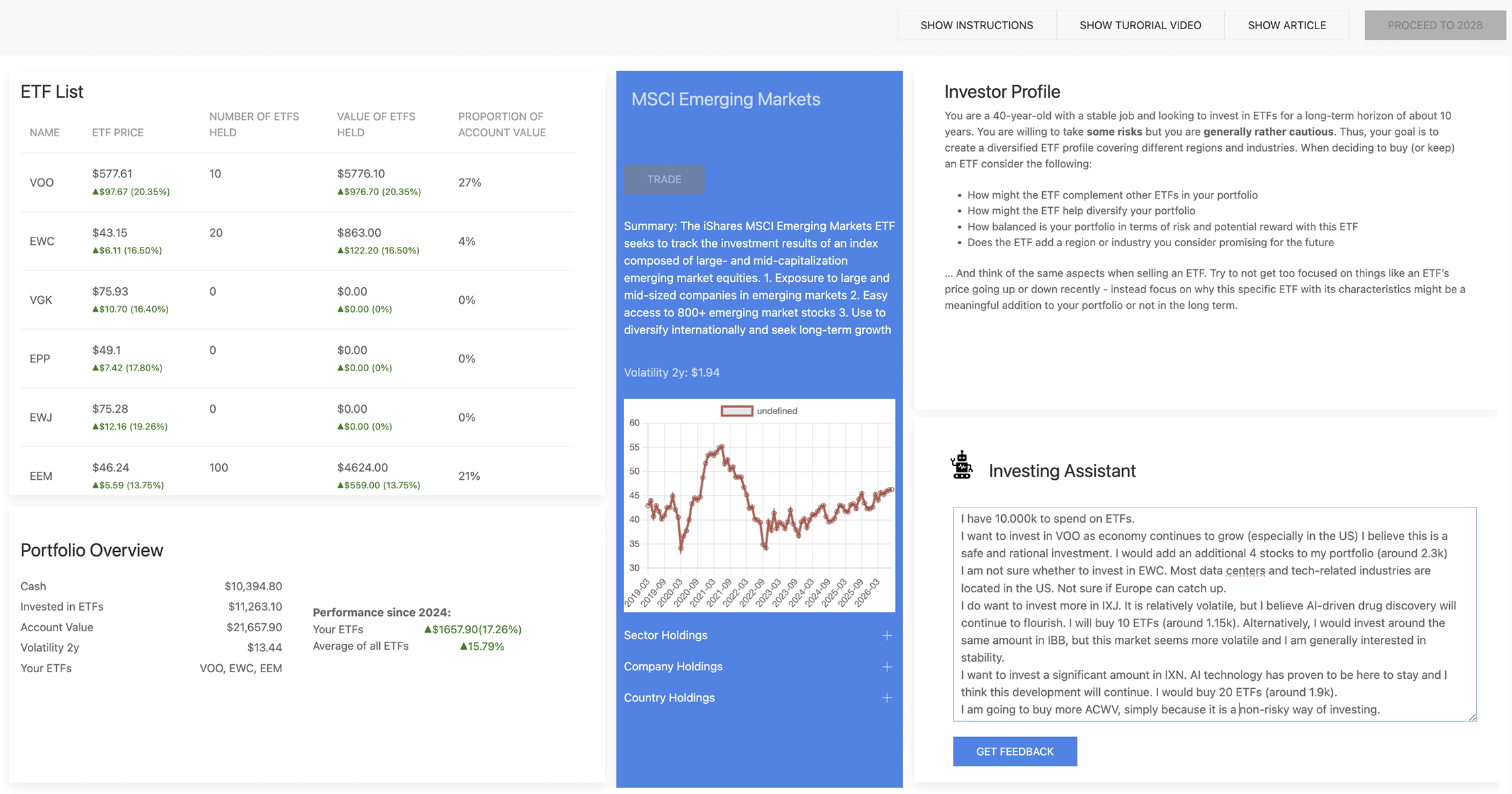}
    \caption{Screenshot of the simulated ETF trading platform interface used in the study, showcasing the various components including the ETF list, portfolio overview, ETF details, investor profile, and the investing assistant.}
    \Description{A screenshot showing the user interface of the ETF trading platform. The interface is divided into several sections: The ETF list, located on the top left, displays ETF names, prices, number of ETFs held, value, and proportion of account value. The portfolio overview, positioned below the ETF list shows cash balance, invested amount, total account value, and portfolio performance metrics. The central panel provides the trade button, as well as detailed information about the selected ETF (in this case, MSCI Emerging Markets). This includes a summary, volatility of the last two years, a historical price chart, as well as sector, company and country holdings of the selected ETF. On the top right, there is the investor profile, outlining the user's investment goals and considerations for ETF selection. The investor profile describes a 40-year-old looking for long-term, diversified investments. At the bottom right, the investing assistant allows users to input their investment rationale and receive feedback. Example data and a rationale are included in the screenshot.}
    \label{fig:trading-platform}
\end{figure*}

For our study, we built a web-based simulated ETF trading platform inspired by existing trading platforms. The application was developed using the Python web framework Flask, JavaScript, and a MySQL database for the ETF data. The main interface of the trading platform consists of three columns: On the left, there is a scrollable list of all available ETFs along with their prices, performance, and details about those already held by a participant (see \autoref{fig:trading-platform}). Below that, there is an overview panel of a participant's portfolio. The middle column displays detailed information about the ETF currently selected in the list, including a short description of how it is composed and what it tracks, its volatility, a price chart over past years and information on sector, company, and country holdings. The right column showed the assistants (depending on the study condition) and a description of the investor profile, which participants should `build' their portfolio for. The screenshot in \autoref{fig:trading-platform} shows a user's typed rationale in the text input field of the \emph{ExtendAI} before sending it out to receive feedback. In the \emph{RecommendAI} condition, no text input field is available, instead, a button labelled ``Get Recommendation'' is available, which returns a recommendation of which ETFs to add or remove from one's portfolio.

Both assistants utilised the OpenAI API using the GPT-4o model. In the API calls for both assistants, the following was included: (1) the user’s current portfolio composition (list of ETFs), (2) a JSON containing detailed information on each available ETF, as described in \autoref{sec:etf-data}, and (3) the investor profile along with the instructions for how the AI should provide its suggestions. The main difference between both assistants (apart from their instructions) is that in  \emph{ExtendAI}, the API call also included the user's rationale which the AI was instructed to extend with its feedback. To provide the feedback, we prompted \emph{ExtendAI} to improve the user's rationale by drawing on its knowledge of the return, volatility, and diversification of each ETF. The feedback is designed to highlight potential blind spots and to provide additional factors to consider, giving the user a more comprehensive basis for their decisions. High-quality feedback would thus add new and non-obvious considerations to the user's rationale, grounded in the information provided to the LLM. The instructions for both assistants were iteratively developed, tested, and refined by four of the authors with various portfolio compositions and, in the case of \emph{ExtendAI}, with different user rationales from the pre-study (\autoref{sec:pre-study}), with the aim to achieve meaningful suggestions from the LLM. \autoref{app:ai-examples} provides some examples for the outputs of both assistants, and \autoref{app:prompts} gives the final metaprompts.

Beyond the general distinction between the metaprompts for both AI roles for providing recommendations versus embedding feedback into the submitted rationale, in \emph{ExtendAI}, we instructed the AI to avoid making specific ETF suggestions. Instead, we focused on highlighting key factors to consider, occasionally referencing specific ETFs as potential alternatives, but primarily emphasising broader aspects for evaluation. The reason for this was that we intentionally wanted to create a contrast between both types of assistance to elicit participants' reflections on the advantages and disadvantages of both. At the same time, we aimed to focus the study on the two distinct roles of the LLM by only offering a one-shot interaction for both assistants, meaning that suggestions from both assistants could only be requested once. Thus, unlike chat-based interfaces such as ChatGPT, where users can engage in multiple exchanges, \emph{ExtendAI} required users to provide a relatively comprehensive description of their rationale upfront.

\section{Evaluation}
In this section, we describe the study we conducted to evaluate the prototype containing the two different types of assistance, \emph{ExtendAI} and \emph{RecommendAI}. The study received ethics approval of \anon[redacted for review]{the IRB at UCL under code UCLIC\_2024\_003\_Rogers}.

\subsection{Participants}
We recruited $N=21$ participants (5 females, 16 males; $M=30.4$ years, $min: 20$ years, $max: 44$ years; see \autoref{tab:participants} in \autoref{app:participants} for details). We initially started recruiting participants through the crowd-sourcing platform Prolific. However, despite using pre-screeners and clearly emphasising in the study description that participants should have experience investing into ETFs, four out of five participants who were recruited could not answer the questions on their experience and knowledge of ETFs at the beginning of the study sufficiently well (e.g., what their strategy is, what they consider when investing, examples of ETFs they have invested in). Consequently, we shifted to recruiting participants through our personal networks. We used our organisations' mailing lists and targeted people directly whom we knew had the required knowledge and/or experience. This allowed us to recruit a sample of participants with the required investing experience and who have a range of different backgrounds (also see \autoref{tab:participants}).
The final sample included $N=18$ participants who had a good understanding of and experience in investing into ETFs (with an average experience of $M=4.47$ years, $SD = 3.88$), and $N=3$ participants with a good understanding of investing and the stock market, but without having first-hand experience in investing in ETFs. In our pre-task questionnaire, participants self-reported their knowledge of the financial markets and of ETFs to be at $Md=4$ and an $IQR=1$ respectively on a 5-point Likert scale. A majority of $N=18$ participants described themselves as long-term investors, in line with the framing of the study task.

\subsection{Study Procedure}
\label{sec:procedure}

\begin{figure*}
    \centering
    \includegraphics[width=0.9\linewidth]{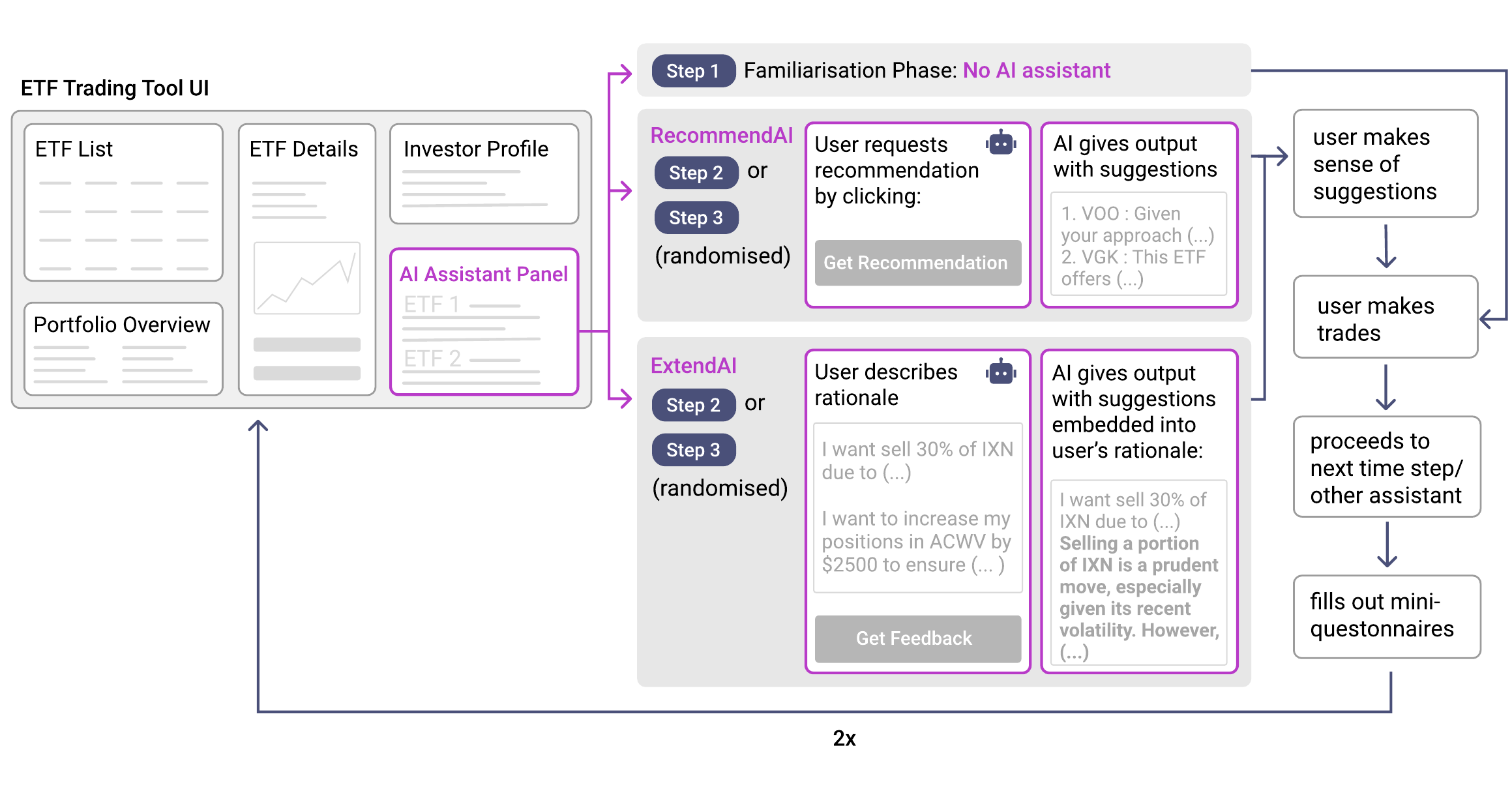}
    \caption{A schematic overview of the ETF trading platform, illustrating what participants did and what the AI assistant panel showed in each of the three steps of the study: familiarisation phase, \emph{RecommendAI}, and \emph{ExtendAI}.}
    \Description{This figure illustrates the ETF trading tool UI, and a detailed user flow, demonstrating the differences between the three study steps. On the left side of the figure, there is a simplified representation of the ETF Trading tool UI. The left panel of the UI shows a list of available ETFs and an overview of their portfolio. The middle panel in the UI displays detailed information about the selected ETF, and on the panel on the right, there is the investor profile at the top and the AI assistant (when available) at the bottom. The right part of the figure shows the study process beginning with the familiarisation phase, where the user interacts with the interface without AI support, makes trades, and then proceeds to the next year. After the familiarisation phase, the user is randomly assigned to either RecommendAI or ExtendAI: a) For RecommendAI, the user requests recommendations by clicking a button. The AI generates and provides output with suggestions (an excerpt of an example suggestion is shown). The user makes sense of these suggestions and makes trading decisions. b) For ExtendAI, the user inputs their investment rationale in a text area and the AI analyses the input and provides suggestions embedded within the rationale (an excerpt of an example input and output is shown). The user reviews the augmented rationale and makes trading decisions. After completing the interaction with each of the assistants the user fills out mini questionnaires.}
    \label{fig:study-flow}
\end{figure*}

The study was conducted over Zoom\footnote{\url{https://zoom.us/}}, following a detailed protocol elaborated through a pilot study with three participants (who were not part of our final study sample). At the beginning of the Zoom call, participants gave their informed consent and filled out a pre-questionnaire covering basic demographics and their investment experience. After the pre-questionnaire, participants were forwarded to our simulated ETF investment platform for the main task, which was to build and revise a portfolio over three time steps: August 2024, August 2026, and August 2028. During the entire study, the Zoom meeting remained open, so that participants could ask clarifying questions on how to perform the task and inform the researcher in case of any issues during the study.

Participants started with 10,000 USD to invest in the first time step. In each of the later time steps, they received an additional 10,000 USD that they could invest in addition to cash earned from selling ETFs. In the first time step, the familiarisation phase, participants had to make their decision without AI assistance. This allowed them to familiarise themselves with the interface of the trading platform and the available ETFs. In each of the later time steps (2026, 2028), participants either used \emph{ExtendAI} or \emph{RecommendAI} during their decision-making (see \autoref{tab:timesteps}). We randomised the order of the two types of assistance across participants. 11 participants used RecommendAI first and 10 used ExtendAI first. See \autoref{fig:study-flow} for more details.

\begin{table}[t]
    \caption{Overview of the high-level task, the form of assistance provided, and which data was shown in the simulated trading interface for each step of the within-subjects study.}  
    \label{tab:timesteps}
    \centering
    \footnotesize %
    \begin{tabular}{p{0.06\columnwidth} p{0.09\columnwidth} p{0.22\columnwidth} p{0.20\columnwidth} p{0.22\columnwidth}}
        \toprule
        \textbf{Step} & \textbf{Year} & \textbf{Task} & \textbf{Assistance} & \textbf{Data Shown} \\
        \midrule
        \end{tabular}
        
        \renewcommand{\arraystretch}{1.5} %
        \begin{tabular}{p{0.06\columnwidth} p{0.09\columnwidth} p{0.22\columnwidth} p{0.20\columnwidth} p{0.22\columnwidth}}
        1 & 2024.08 & 
        \parbox[t]{0.22\columnwidth}{\raggedright Participant makes trades to \textit{build} initial portfolio} & 
        \parbox[t]{0.20\columnwidth}{\raggedright No AI assistant (familiarisation phase)} & 
        \parbox[t]{0.22\columnwidth}{\raggedright ETF price data and econ. narrative based on \textit{real data}} \\

        2, 3 & 2026.08, 2028.08 & 
        \parbox[t]{0.22\columnwidth}{\raggedright Participant makes trades to \textit{revise} portfolio} & 
        \parbox[t]{0.20\columnwidth}{\raggedright RecommendAI or ExtendAI in random order} & 
        \parbox[t]{0.22\columnwidth}{\raggedright \textit{Synthetic} ETF price data and econ. narrative} \\

        4 & 2030.08 & 
        \parbox[t]{0.22\columnwidth}{\raggedright Participant checks final portfolio value} & 
        \parbox[t]{0.20\columnwidth}{\raggedright N/A} & 
        \parbox[t]{0.22\columnwidth}{\raggedright Final portfolio value (based on synthetic data)} \\
        \bottomrule
    \end{tabular}
\end{table}

At the beginning of each time step, participants watched a short tutorial video about the trading platform (first time step, 2024) or the respective AI assistant (second and third time steps, 2026/2028) and were presented with the `news article' (\autoref{sec:synthesising-data}) covering relevant economic and market-related events as background information.
After each of the three time steps, participants filled out a short questionnaire about the decision they made.
After each of the later two time steps with AI assistance (2026, 2028), participants were asked a short set of questions about their experience of using the respective AI assistant and how they perceived its impact on their decision-making 

We chose not to collect think-aloud protocols since the task was already rather demanding on its own and required participants to be fully focused. After the main task, we conducted exit interviews of $M = 23$ min ($SD=6.8$ min), focusing on participants' decision-making process and how they thought the two types of assistance affected it. The full interview guide is provided in the supplementary material. We recorded audio through Zoom only for the interviews. In total, the study lasted $M = 91$ min ($SD=17.5$ min), depending on how fast participants finished the main task. Participants were compensated with Amazon vouchers worth 30 EUR or an equivalent amount in participants' preferred currency.

\subsection{Measures}
A set of measures were used covering the time spent on the different study parts, the number of interactions with the trading platform, and the trades participants made. Furthermore, measures covered (1) participants' assessments of their decisions, (2) their experience with the two types of assistance, and (3) the impact they had on their decision-making, described in more detail in this section.

As part of the study task, participants filled out a brief questionnaire after each of the three steps, in which they were asked to report on five-point Likert scales how well informed they felt about their decision, how confident they were in it, and how satisfied they were with it. After completing these items, participants were shown the performance of their portfolio after two years and were asked to re-evaluate the confidence and satisfaction items. The first two questions were asked before revealing how the portfolio developed over the two years following the current time step, while the third question was asked after revealing the outcome. This perceptions-based approach was informed by \citet{bradburyImprovingInvestmentDecisions2014} and serves as a proxy for decision quality. The reasoning is that good financial decisions are marked by how well people understand the risks of their decisions.

At the end of the second and third time step (2026, 2028), after having completed the set of trades with the help of one of the two assistants, participants were asked a short set of questions about its helpfulness, perceived impact on their decision-making process (including cognitive load using a shortened version of NASA-TLX~\cite{hartDevelopmentNASATLXTask1988}), and an item measuring participants' sense of agency using a slider from -50 (``decision was fully made by AI'') to +50 (``decision was fully made by myself''). The complete questionnaires with all the items can be found in the supplementary materials.

To measure the impact the two types of assistance had on the decision quality, we measured how the portfolio diversification changed at each time step. This was done through four metrics: (1) the number of countries included in the portfolio, (2) how balanced the allocation was across these countries, and (3) how balanced the allocation was across the 11 sectors (based on the Global Industry Classification Standard).

We further measured the magnitude of the AI assistants' impact on participants' decisions. The impact is conceptualised differently for both assistants, given the different decision-making steps they involve. For \emph{ExtendAI}, participants first developed a plan, hence, the impact was measured by the number of ETFs in their final trades that were \emph{not} part of their initial plan. For \emph{RecommendAI}, which provided recommendations without a prior plan, the impact was measured by the number of trades in the final list that were directly suggested by the AI. These metrics serve as proxy measures of the AI's impact and are used exploratively to examine the interactions with the tools and the two AI assistants.

\subsection{Data Analysis}
Our \textit{quantitative analysis} relied on data collected in the questionnaire before the study tasks and three questionnaires during the study task.
In the analyses, we only report descriptive statistics and refrained from inference testing, given the exploratory nature of our study. To refer to central tendency and spread, we use the mean ($M$) and standard deviation ($SD$) for continuous variables, and median ($Md$) and interquartile range ($IQR$) for ordinal variables. 

For our \textit{qualitative analysis}, all interview recordings were transcribed verbatim using Zoom's automatic transcription, followed by manual cleaning by one researcher, and subsequently imported into Atlas.ti Web\footnote{\url{https://web.atlasti.com/}}. Initially, four researchers conducted open coding on a subset of eight interviews (four interviews each, with allocations offset so that one researcher shared two interviews with one, and two with another researcher). A coding tree was iteratively developed over four meetings between the four researchers. The interviews were then distributed among three of the researchers for coding using the predefined coding tree. In two meetings among the three researchers, a reflexive thematic analysis was conducted to identify overarching themes~\cite{braun2023TApractice,braun2006thematic}.

\section{Findings}
We gathered both quantitative data from the questionnaires and interactions with the prototype, as well as qualitative data from the interviews, which we present in this section.

\begin{figure*}[htbp]
    \centering
    \begin{minipage}[b]{0.45\textwidth}
    
        \centering
        \includegraphics[width=0.85\textwidth]{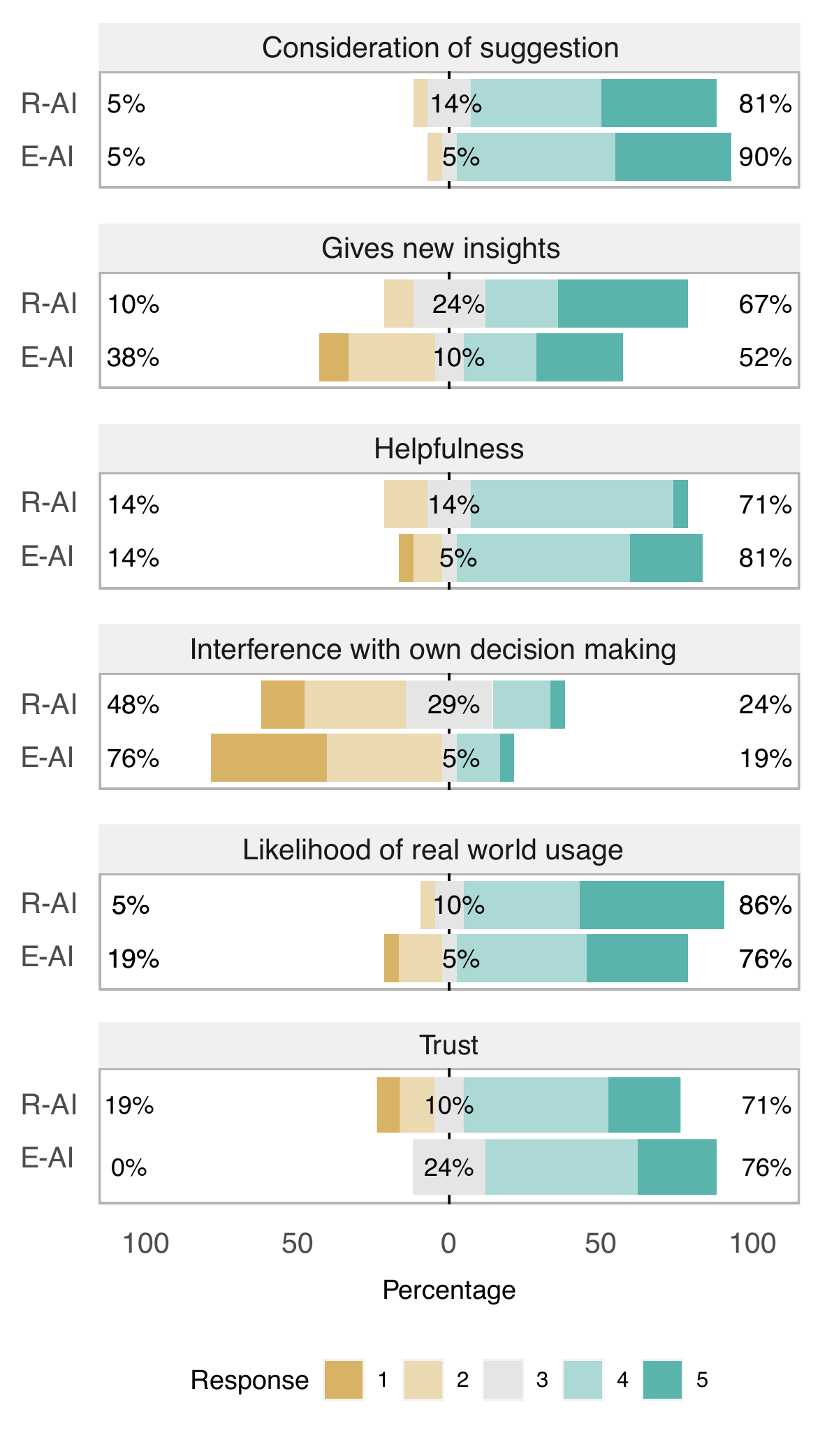}
        \caption{Participants' scores on a 5-point scale for the questions on AI suggestion consideration, insights, helpfulness, interference with decision-making, likelihood of usage, and trust. R-AI refers to \emph{RecommendAI} and E-AI to \emph{ExtendAI}.}
        \Description{The figure shows the distribution of Likert scores of six of the questionnaire items in percent. The plot shows how many percent of participants chose which of the five options from 1 (disagree strongly) to 5 (agree strongly). The first line shows the responses to the question if participants considered the AI's suggestion in their decision-making to which 81\% in RecommendAI agreed and 90\% in ExtendAI. The second row shows the answers to the item if the AI gives new insights, which 67\% in RecommendAI agreed and 52\% in ExtendAI agreed to. The third row shows the ratings of the AI's helpfulness, receiving 71\% in RecommendAI agreed and 81\% in ExtendAI. The fourth row shows the ratings to the item if the AI interferes with one's decision-making process which was rated with 24\% in RecommendAI 19\% in ExtendAI, indicating participants perceived higher interference from RecommendAI. The fifth row shows participants' evaluation of the likelihood of real-world usage of the AI, where RecommendAI got 86\% and ExtendAI 76\%. The sixth row shows users trust in both AIs where 71\% stated that they trust R-AI and 76\% that they trust ExtendAI.}
        \label{fig:image1}

        \vspace{4em}

        \centering
        \includegraphics[width=0.90\textwidth]{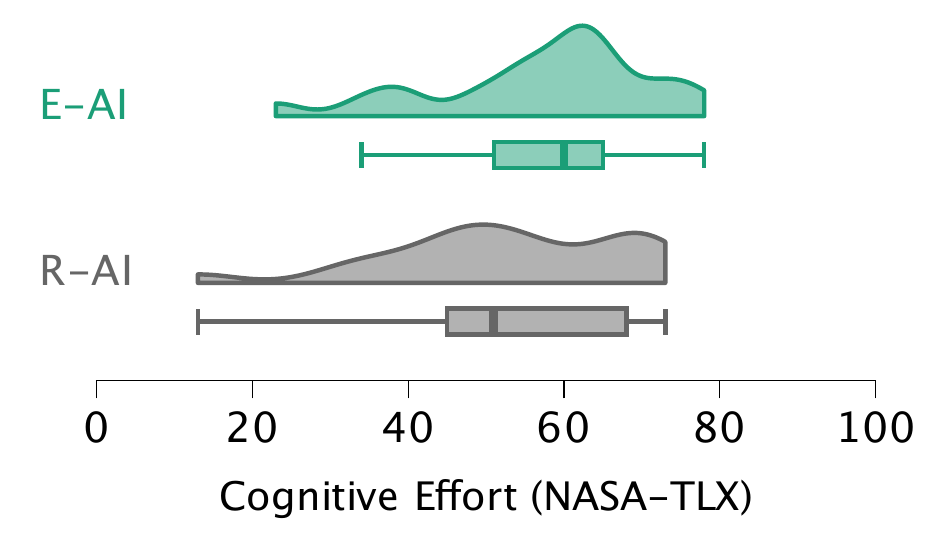}
        \caption{Distribution of Nasa-TLX raw scores indicating cognitive effort for \emph{ExtendAI} (E-AI) and \emph{RecommendAI} (R-AI).}
        \Description{The figure shows a `raincloud' plot and a box plot representing the distribution of scores of the NASA-TLX for both AI assistants. It can be seen in the plots that ExtendAI generally received higher ratings than RecommendAI.}
        \label{fig:image4}
        \vspace{5em}
    \end{minipage}
    \hfill
    \begin{minipage}[b]{0.45\textwidth}

        \centering
        \includegraphics[width=0.90\textwidth]{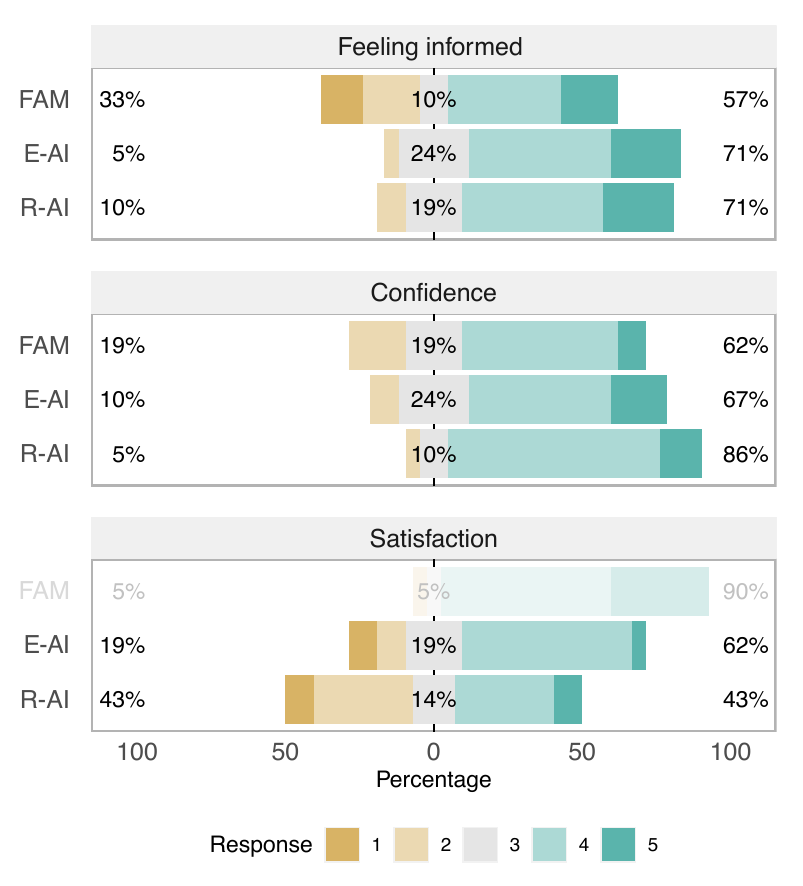}
        \caption{Participants' scores on a 5-point scale on the extent of feeling informed, their confidence in their decisions, and their satisfaction with decision outcomes for the familiarisation phase (FAM) \textit{RecommendAI} (R-AI) and \textit{ExtendAI} (E-AI). \textit{FAM} is greyed out for \textit{Satisfaction}, as it cannot be meaningfully interpreted.}
        \Description{The figure shows the distribution of Likert scores of the items referring to participants' assessments of their decisions with respect to how well-informed they felt (first Likert plot), their confidence in their decision (second Likert plot), and satisfaction with their decision outcome (third Likert plot).}
        \label{fig:decisionassessment}

        \vspace{3em}
        
        \centering
        \includegraphics[width=0.85\textwidth]{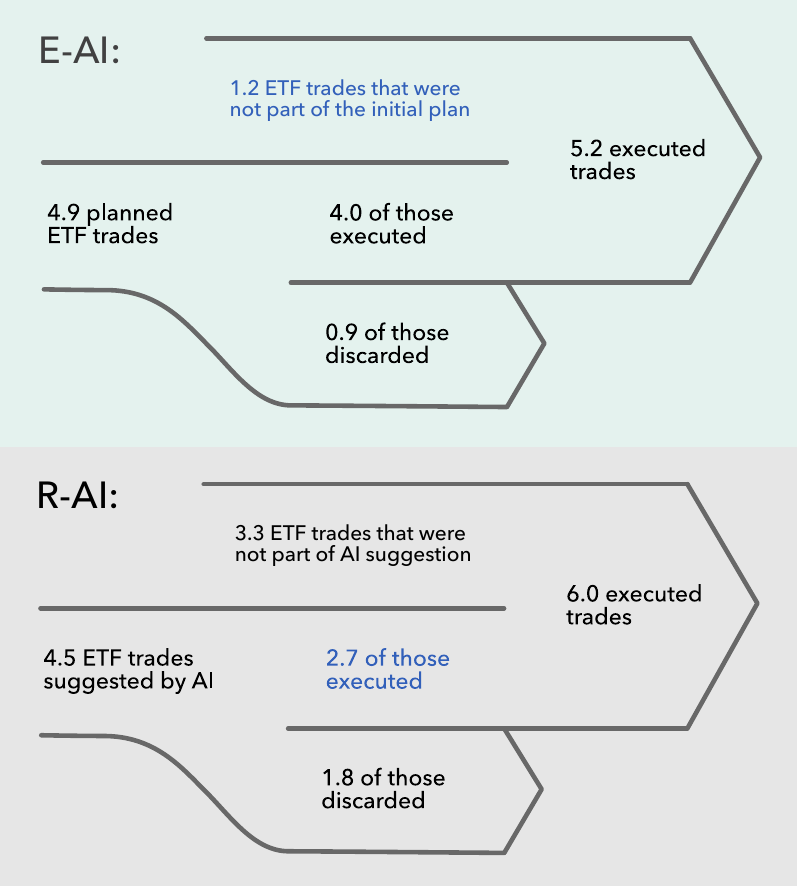}
        \caption{A depiction of the `impact' of AI assistants on decision-making (reads from left to right) with blue elements indicating the AI assistants' impact: for \emph{ExtendAI}, it is calculated as the percentage of trades ultimately made that were not part of a participant's initial plan (i.e 1.2 trades / 5.2 trades = 23.08\%), for \emph{RecommendAI}, it is calculated as the number of trades made which were directly suggested by the AI (i.e. 2.7 trades / 6 trades = 45.00\%).}
        \Description{A visual representation of the `flow' of decision-making for both AI assistants using averages. For ExtendAI, the flow starts with the number of planned trades (4.9), which then splits into which of these trades are executed (4.0) or discarded (0.9) by the user. To those that are executed an additional (1.2) are added which are from those suggested by the AI resulting in an average of 5.2 trades finally being made. These 1.2 additional trades indicate the impact ExtendAI had on the participant's final decision outcome. For RecommendAI a similar representation is provided. However, in contrast to ExtendAI it starts with the trades suggested by the AI (4.5) which then splits into which of these trades are executed (2.7) and which of them are discarded (1.8). To those that are executed additional trades which the participant came up with independent of the AI are added (3.3) resulting in an average of 6.0 trades finally being made.}
        \label{fig:image2}
    \end{minipage}

\end{figure*}

\subsection{Quantitative Findings} \label{sec:quant}
\subsubsection{Task Duration} Overall, participants spent $M=44.00$ min ($SD=13.28$) on the investing task. In Step~1, participants traded ETFs without assistance (familiarisation phase), in Step~2 and 3, they interacted with either \emph{ExtendAI} or \emph{RecommendAI} (order randomised). Participants spent $M=17.99$ min ($SD=6.14$) in the familiarisation phase, $M=17.46$ min ($SD=6.03$) on \emph{ExtendAI}, and $M=8.56$ min ($SD=3.37$) on \emph{RecommendAI}. For \emph{ExtendAI}, participants spent $M=8.38$ minutes ($SD=4.69$) on writing their rationale. After participants received the suggestions from either assistant, participants started trading after $M=1.75$ min ($SD=1.16$) with \emph{ExtendAI}, and after $M=2.08$ min ($SD=1.46$) with with \emph{RecommendAI}.

\subsubsection{Interactions With Trading Tool} Participants looked at $M=28.86$ ($SD=4.91$) of the available 31 ETFs in the familiarisation phase. This was measured by counting the ETFs that were clicked on to display more information. In \emph{ExtendAI}, participants  looked at $M=19.67$ ($SD=9.23$) of the ETFs, and in \emph{RecommendAI} at $M=12.86$ ($SD=8.79$).

\subsubsection{Overall Development of Portfolio} Participants started off with a portfolio containing $M=5.48$ ($SD=4.07$) ETFs in the familiarisation phase (year 2024). The number of ETFs in participants' portfolios then tended to increase over the following two time steps to $M=7.48$ ($SD=4.30$) in the second time step (2026), and to $M=8.86$ ($SD=4.48$) in the third time step (2028).

\subsubsection{AI Assistants' Impact on Portfolio Composition}
The set of trades participants made following an interaction with one of the assistants generally tended to result in an increase in the number of countries held. In \emph{ExtendAI}, the increase was $M=3.00$ ($SD=5.933$) and in \emph{RecommendAI} slightly smaller $M=2.33$ ($SD=4.247$). Furthermore, the deviation of an equal distribution across these countries also decreased for both assistants. The decrease of this deviation tended to be more pronounced in \emph{ExtendAI} with $M=-0.732$ ($SD=1.211$) than in \emph{RecommendAI} where it was $M=-0.369$ ($SD=1.135$). The distribution across sectors was generally also improved following the interaction with both assistants. However, here the decrease in the deviation in \emph{ExtendAI} $M=-0.275$ ($SD=1.741$) tended to be slightly smaller than in \emph{RecommendAI} $M=-0.349$ ($SD=1.498$).

\subsubsection{AI Assistants' Influence on Decision-Making} Participants made $M=5.19$ ($SD=2.40$) trades with \emph{ExtendAI}, and $M=5.95$ ($SD=3.46$) with \emph{RecommendAI}. In \emph{ExtendAI}, 23.08\% of the ETF investment decisions participants made were different to what they described in their rationale before interacting with the assistant. In \emph{RecommendAI}, 45.00\% of their decisions were based on the assistant's recommendation (depicted in \autoref{fig:image2}). These percentages are proxy measures for the assistants' impact on participants' decision-making.

\subsubsection{Perceptions and Assessment of Decisions}
In this and the following subsection, we will only report participants' \textit{agreement} to the separate items for readability and simplicity (i.e., how many participants gave answers above the neutral point). Furthermore, we only report on those items with meaningful differences between the two assistants. For a more detailed picture of how participants answered these questions, see \autoref{fig:image1} and \autoref{fig:decisionassessment}. In terms of how well-informed participants considered their decisions to be, in the familiarisation phase, 57\% stated that they felt informed, whereas with assistance, this was the case for 71\% (for both assistants). With respect to their confidence in their decision, the familiarisation phase reached 62\%, and participants felt more confident with \emph{RecommendAI} (86\%) than with \emph{ExtendAI} (67\%). However, in terms of how satisfied participants were, the ratings for the assistants were `flipped' with \emph{RecommendAI} only reaching 43\% and \emph{ExtendAI} reaching 67\%. Note that a comparison with the familiarisation phase is not meaningful/adequate with respect to satisfaction, since in the familiarisation phase participants' portfolios generally performed better than in the two following time steps when using the assistants (the synthetic ETF prices were intentionally created this way, as described earlier). 

\subsubsection{Perceptions and Experiences With Both Assistants} As can be seen in \autoref{fig:image1}, participants reported that they generally trusted both assistants---\emph{ExtendAI} with 76\% receiving a slightly higher rating than \emph{RecommendAI} with 71\%. Furthermore, participants reported that they considered the assistants' input (81\% \emph{RecommendAI} and 90\% \emph{ExtendAI}). For \emph{RecommendAI}, participants generally found that it gives them new insights (67\%), while for \emph{ExtendAI} this was less the case (52\%). Despite this somewhat lower rating for insights, 81\% of participants considered \emph{ExtendAI} to be helpful, whereas only 71\% thought this was true for \emph{RecommendAI}. Participants generally did not feel that the assistants interfered with their decision-making process, although an even smaller number of them thought this to be the case for \emph{ExtendAI} (19\%) than for \emph{RecommendAI} (24\%). Participants reported that they would use an assistant like \emph{RecommendAI} (86\%) in real-world settings slightly more than \emph{ExtendAI} (76\%). This might be related to the aforementioned finding that participants felt like they got more insights from \emph{RecommendAI}, as well as the fact that they rated it to be less cognitively demanding (see \autoref{fig:image4}), where \emph{ExtendAI} received a NASA-TLX Raw score of $M=57.00$ ($SD=14.02$), and \emph{RecommendAI} $M=52.52$ ($SD=15.55$).

\subsection{Qualitative Findings}
\label{sec:qual}
As described in \autoref{sec:use_case}, we designed both assistants to trigger different decision-making processes, which was also reflected in how participants described their experience of using each assistant. One participant, who summed it up well, described his process with \emph{RecommendAI} as: ``\textit{I first look into the recommendations. Then I decide for myself whether it makes sense or not.}'' For \emph{ExtendAI}, the same participant described his process as follows: ``\textit{The first part of the decision-making was exactly identical [to the first step without any assistant], so write down everything and ask for feedback. And then reconsider it if there's something significant that I've missed.}'' 

Participants' opinions of these two types of assistance differed greatly, with 11 participants stating in the interview that they preferred RecommendAI, and 10 participants preferring ExtendAI. We identified four themes in our thematic analysis that explore these differing opinions from various angles. In the following, we denote participants who first interacted with RecommendAI and then with ExtendAI as RE-$x$, and participants with the reverse order as ER-$x$. Participants frequently referred to the assistants using variations of `the first one' or `the second one'. To improve readability, we replaced these with \textit{[RecommendAI]} or \textit{[ExtendAI]} in the quotes.

\subsubsection{\textbf{Theme 1:} Help Me, but Let Me Think for Myself}
\label{sec:theme_1}
As mentioned above, participants' preferences for the two types of assistance were evenly split. On the one hand, some participants preferred \emph{RecommendAI} as they considered it easier to use; not only because ``\textit{it's more convenient to just press the button and get a result}'' (ER-8), but also because these participants considered the concrete ETF recommendations more actionable:
\begin{quotation}
    ``\textit{It's highly comparable to a friend who is deep into finance, and he tells me, I've heard about this health care stuff, maybe have a look at it, [...] and I kind of trust it. And then I basically quickly reevaluate it and then directly go for it.}''~(RE-7)\\
    ``\textit{[RecommendAI] gave me some hints, which made it a bit quicker to make decisions.}''~(RE-8)
\end{quotation}
Conversely, some participants rejected the immediate shift into specific recommendations, because they felt excluded from the decision-making process:
\begin{quotation}
    ``\textit{It was far less engaging and it did not know what I want to do at this stage. It was recommending stuff. [...] It doesn't know my current state of mind [...].}''~(ER-1)\\
    ``\textit{My main issue with [RecommendAI] was that it directly gave me some kind of `do this, then do this, then do this,' which I, in some sense, followed without thinking too much about it. So, I gave the AI assistant more freedom than I usually would have and thought about it even a bit less.}''~(ER-9)
\end{quotation}
These participants preferred \emph{ExtendAI}'s more restrained approach, noting that it gives them more agency and freedom to think for themselves:
\begin{quotation}
    ``\textit{I kinda liked the style that it was having this somewhat passive voice, and not aggressively recommending you to buy, buy, buy, or sell, sell, sell something.}''~(ER-7)\\
    ``\textit{I think it gives me a bit more control and agency. [...] It didn't just tell me, `This is what you should do.' [...] It kind of supports how I'm thinking [...], it gives me information to help me decide.}''~(RE-12)
\end{quotation}
This helped participants to ``\textit{think more over my decisions}''~(RE-11) and to ``\textit{really make you reflect}''~(ER-8). However, other participants found \emph{ExtendAI}'s feedback too general and not actionable enough. For instance, one participant stated that ``\textit{it just gave me some comments on the ETFs I selected. But I didn't know, how can I start with the feedback, and what should I do instead? So it was not really helpful}''~(ER-2).

These varying perspectives from participants highlight a tension between actionability and helpfulness of AI suggestions on the one hand, and preserving users' agency and their own reasoning on the other hand. The more specific the AI suggestions are, the easier it is to act on them, but the less room it leaves for users' own reasoning, and vice versa. While this trade-off reflects the general tendency in participants' opinions, it is important to note that some participants had the exact opposite experience. For example, one participant disagreed that \emph{RecommendAI} reduced his own reasoning, emphasising that ``\textit{I was still reflecting. I was probably reflecting even more with \textit{[RecommendAI]}}''~(ER-4). Another participant on the other hand rejected the common view that \emph{RecommendAI} was easier to act on, explaining that ``\textit{it was actually more work because it had a lot of information. I wanted to check every point of those. Is it true or not? And is the logic making sense or not?}''~(ER-1). This participant found that \emph{ExtendAI} only required a ``\textit{small check}''~(ER-1), since its suggestion built upon his own reasoning making it easier for them to make sense of what it said, which leads into the next theme.

\subsubsection{\textbf{Theme 2:} Insights Should Be New, but Aligned With the User's Thinking}
Both assistants tended to add to participants' reasoning in different ways. \emph{RecommendAI} tended to be more helpful for discovering new ETFs that participants did not consider before, or more generally, for thinking into new directions, as its suggestions were considered more ``\textit{out of the box}''~(RE-4), making it more likely ``\textit{to think about things you didn't think of before}''~(RE-9). \emph{ExtendAI}, on the other hand, tended to help participants think through their own reasoning more thoroughly and revealed blind spots in their rationale:
\begin{quotation}
    ``\textit{So when I said, `Oh, I’m thinking about investing into, I think it was IBB and IXJ, two biotech-related ones', and my initial idea was to just invest in one. And it was like, `Well, why don't you diversify because they're differently focused?' That was a good suggestion that I hadn't thought about before.}''~(ER-10)
\end{quotation}

Yet, both assistants had their drawbacks. Some participants found \emph{RecommendAI}'s suggestions challenging to verify due to their unclear basis, describing them as ``\textit{black-box solution coming from nowhere}''~(ER-1), ``\textit{based on whatever analysis it does, we don't know anything about it}''~(ER-3). This was in contrast to \emph{ExtendAI}, whose feedback some participants found easier to verify as it was integrated into their own reasoning:
\begin{quotation}
    ``\textit{It allows me to delve a bit deeper into the problem and understand the recommendations from the AI assistant a bit more than in [RecommendAI], where I somewhat blindly trust it. If I first have to gather the information by myself and evaluate it, then I think I’m better informed to also evaluate the truthfulness of whatever the AI assistant recommends.}''~(ER-9)
    \\``\textit{I've read the feedback more carefully. Let's say, if it would have been longer, I would have still easily read it. Whereas in [RecommendAI] you sometimes go a bit in this skim-reading mode, that you read the headline and then you conclude something.}''~(ER-7)
\end{quotation}
However, \emph{ExtendAI} building upon participants' rationale also frequently led to a perceived lack of novel input to their reasoning:
\begin{quotation}
    ``\textit{I feel like [ExtendAI] would be more useful if it was less stuck in my way of doing things.}''~(RE-9)\\
    ``\textit{I felt like these were just direct feedback on my flawed suggestions. And I wish that I could have a fresher input, actually.}''~(RE-10)
\end{quotation}

These complementary strengths and weaknesses of the two assistants underscore another tension: The insights provided by AI should be `fresh' and new to be of value to the user. At the same time, these insights need to be consistent with the user's reasoning so that the user accepts them and can more easily integrate them into their decision-making. The challenge of navigating this tension is further illustrated by participants' statements about their reliance behaviour. With both assistants, participants tended to ignore AI suggestions that were not aligned with their thinking. This usually became apparent for participants who referred heavily to the economic `news articles' in an attempt to anticipate the short-term performance of individual ETFs, and expected the assistants to help them react to the developments and events described in the news articles. However, recall that the assistants were---consistent with the investor profile provided---designed to consider only the diversification and long-term potential of the ETFs, not the recent and expected short-term performance of the ETFs, as explained in \autoref{sec:interface-and-ai-design}. As a result, these participants frequently found \emph{ExtendAI}'s feedback to be ``\textit{too general, too plain}''~(ER-2), feeling that ``\textit{it didn't really give much insight that changed my mind}''~(RE-2), and thereby disregarding \emph{ExtendAI}'s comments on how to improve the diversification of their portfolios. For \emph{RecommendAI}, these participants ignored its recommendations when they found them to contradict the economic news articles, even though the recommendations may be sensible from a long-term perspective: ``\textit{I completely ignored this whole S\&P 500. The whole thing is going down, it looks like it is not the time to invest. Everyone is selling}''~(ER-1).

Conversely, we also observed the opposite effect to be possible for both assistants, as in other cases the AI suggestions appeared to be \textit{too} closely aligned with the participants' thoughts. Consequently, these participants frequently found that \emph{ExtendAI} gave ``\textit{confirmation that you have the right thoughts''}~(RE-5), or ``\textit{reassures me that I'm not doing something wrong}''~(RE-8), rather than adding new insights or highlighting weaknesses in the portfolio. This perception might have been strengthened by participants' tendency to focus on parts of the feedback that reinforce their reasoning, as one participant observed: ``\textit{If I reflect now, I was kind of looking for sentences confirming my strategy.}''~(RE-7). We observed this also with \emph{RecommendAI}, albeit less frequently: ``\textit{For the first AI assistant, most of the recommendations I also had thought about before. And so, it was like a confirmation of my reasoning}''~(RE-11). While these confirmations meant participants ``\textit{felt more confident in taking the investment decisions}''~(RE-5), it could also lead to unwarranted confidence, as pointed out by one participant: ``\textit{But the thing is, if the AI assistant has confirmed my thought, then I will be overconfident}''~(RE-4). 

Overall, this theme highlights the challenge of simultaneously aligning with users' thinking \textit{and} helping them arrive at new thoughts. The next theme explores when AI is most likely to be able to extend users' thinking in this way.

\subsubsection{\textbf{Theme 3:} AI Is Most Insightful When Human Thoughts Are `In the Making'}
\label{sec:theme_3}
As described at the beginning of \autoref{sec:qual}, \emph{RecommendAI} usually served as a starting point for participants' decision-making, while \emph{ExtendAI} came in at later stages of the decision-making process. Allowing participants to receive feedback from \emph{ExtendAI} only once, encouraged them to think thoroughly about their decisions before interacting with the assistant; but, in principle, \emph{ExtendAI}'s design left it up to the user how much they wanted work out their rationale before asking the assistant for feedback. Consequently, the rationales that participants provided to \emph{ExtendAI} varied in their level of elaboration (e.g. some rationales included both general strategies, and specific ETFs and sectors, whereas others only focused on specific ETF choices). This  appeared to affect how insightful participants perceived the feedback of \emph{ExtendAI} to be. Participants who were still in the process of making up their minds, or whose rationale was in a less mature state when they interacted with \emph{ExtendAI}, reported that they benefited from the assistant's feedback:
\begin{quotation}
    ``\textit{I was a bit insecure, if I should invest in all these three emerging sectors. But then, when it was really so clear feedback from the AI, I felt like, I had it already in mind a bit like this. But I see it now even more clearly.}''~(ER-5)
\end{quotation}

In contrast, participants who had already invested a lot of thought to fully develop their rationale reported minimal benefit from \emph{ExtendAI}'s feedback. In some cases, this was because the participant's rationale was so sophisticated that \emph{ExtendAI} was not able to point out further improvements:
\begin{quotation}
    ``\textit{I had the feeling, the focus was more on confirming what you're doing, kind of saying yes to what I was doing, rather than the other way around, kind of destroying what I want to do. But from my feeling, my strategy is also not complete bullsh**, so it might be different if the result would be total mess.}''~(RE-7)
\end{quotation}
However, another factor is that participants who have already put considerable effort into their rationale tend to hold on to it, as some participants also admitted upon reflection:
\begin{quotation}
    ``\textit{You have strong belief in what you want to do, right? And at that time you will consider it as noise, this unuseful information. [...] But to be very objective, it might not be noise. But the problem is, you are super subjective, you just want to believe in your belief.}''~(RE-3)
\end{quotation}

We do not see this issue with \emph{RecommendAI}, mainly because its recommendations came before participants had thought deeply about their decision:
\begin{quotation}
    ``\textit{Whereas in [RecommendAI], I haven't taken any decisions. And then I was reading through what the AI assistant would tell me to do, and I was trying to make my mind based from what the assistant is telling me.}''~(ER-4)
\end{quotation}
On the downside, \emph{RecommendAI}'s suggestions were often perceived to come too early in the decision-making process, thereby taking away the opportunity for independent thinking from participants, as outlined in \autoref{sec:theme_1}.  Together, these statements in this theme point out a third tension in augmenting human decision-making: AI suggestions should be introduced at the right moment. Suggestions should not be introduced too early in the decision process to avoid unwanted anchoring around the AI, but also not too delayed to meaningfully contribute to the user's reasoning. This also relates to participants' opinions on using natural language to describe their rationale in \emph{ExtendAI}, as discussed in the next and final theme.

\subsubsection{\textbf{Theme 4:} Natural Language: Flexible Expression and (Un)desirable Friction}
\emph{ExtendAI} relied on users to articulate their reasoning in natural language, which participants felt conflicted about. On the one hand, participants found it ``\textit{just annoying}''~(RE-8) or ``\textit{somewhat cumbersome}''~(ER-7) to type out their entire rationale. This also meant that participants were particularly disappointed when \emph{ExtendAI}'s feedback turned out not useful to them:
\begin{quotation}
    ``\textit{I had to read my data, and then give the input, say, `Hey, I want to decrease this, and I want to do that and that.' And then I only got a comment on this, it was not a new idea on how to invest, or even some numbers. It was just a comment, that it's a good idea.}''~(RE-6)
\end{quotation}
On the other hand, some participants acknowledged the benefit of being forced to consciously think through their decision, ``\textit{to not randomly invest in things, because of simple feelings}''~(RE-7):
\begin{quotation}
    ``\textit{I think if the goal is understanding, then you do have to put in some work yourself. There’s no pain, no gain, I would say.}''~(ER-10)\\
    ``\textit{Yeah, no free lunch, right? So you have to do some work.}''~(ER-1)
\end{quotation}

However, it was not only the effort of typing out or thinking through the decision that participants discussed. Some participants also emphasised the more fundamental difficulty of expressing vague thoughts in words, stating that they ``\textit{had a hard time describing why I'm doing things}''~(RE-7), or that they found it ``\textit{a bit hard to put my strategy into words}''~(RE-8). Then again, some participants also stated that natural language gives them the flexibility to describe their reasoning and complex dependencies in a way that might be difficult to do otherwise:
\begin{quotation}
    ``\textit{I think, especially for [...] complex investment strategies where you take percentages and combine things, [...] maybe combine topics and have some dependencies between emerging trends in the healthcare sector [...]. This is sometimes better to capture in words.}''~(RE-7)
\end{quotation}

All of these comments show that the natural language capabilities of LLMs can be valuable for decision-making, but also a challenge, highlighting the need to employ them purposefully in the design of AI assistants. To this end, participants made several suggestions. For instance, some participants suggested how the writing of the rationale could be simplified by suggesting what to write, or by describing the rationale for one ETF at a time:
\begin{quotation}
    ``\textit{Not have this big, free text field where you might miss [something], because I needed to check from left to right, `Oh, did I mention all the trades, which I've already done, and where I want to do something? Or did I miss one?' So, having kind of, `Okay, here for emerging markets,' you have this text field, where you either write or you have some support, what you could write.}''~(RE-7)\\
    ``\textit{I would do it actually ETF by ETF. [...] Asking me to write a whole strategy, `I want to do this and this, and this and this, and these are the reasoning, blah blah.' This is too much, and I think it will be a huge text. But to have it more specific for a particular ETF, this is helpful. So, `This is a particular ETF, my train of thought is this, and that's what I want to do.'}'' (ER-1)
\end{quotation}
Other participants proposed relying more on visuals and direct manipulation GUIs, and limiting the use of natural language to where it is actually valuable:
\begin{quotation}
    ``\textit{I think integrating visual feedback with the AI's inputs is just more helpful because that's how people are used to making investing decisions.}''~(ER-3)\\
    ``\textit{It would be better if you could make a `pseudo trade', that you have an interface to do a prototype of your trades, add one line of reasoning behind each trade, and then it'll give you feedback.}''~(ER-7)
\end{quotation}

\section{Discussion}
As our findings showed, both types of AI assistance impacted participants in their decision-making and the outcomes they achieved in rather distinct ways, which we further discuss in this section.

\subsection{The Approach of Extending Users’ Own Reasoning Is (Too) Subtle, but Effective}
Both \textit{ExtendAI} and \textit{RecommendAI} helped participants to build more diversified portfolios, with \textit{ExtendAI} being slightly more effective especially in terms of regional diversification. Interestingly, participants achieved this with fewer, but apparently more targeted trades with \textit{ExtendAI} compared to \textit{RecommendAI}, as the analysis of the impact of both assistants on participants' decision-making showed. Combined with participants' statements on how they used both assistants, this suggests that participants understood the weaknesses of their portfolio better when using \textit{ExtendAI}, while they tended to review \textit{RecommendAI}'s suggestions individually, with a less holistic perspective on their portfolio.

Both assistants also made participants feel more informed. %
This perception appeared more warranted in the case of \textit{ExtendAI}, as participants' confidence before seeing the outcome generally aligned with their satisfaction afterwards, suggesting an adequate understanding of the associated risks.
With \textit{RecommendAI} on the other hand, there was a big discrepancy between participants' high confidence before and low satisfaction after seeing the outcome. This can be interpreted as a form of overreliance, similar to the strategy-graded notion of reliance introduced by \citet{fokSearchVerifiabilityExplanations2024}. Another interpretation is that participants perceived less ownership over their decision with \textit{RecommendAI}, similar to how users do not perceive ownership over AI-generated text~\cite{draxlerAIGhostwriterEffect2024}. As a result, participants seemed to be satisfied when the outcome was good and blamed \textit{RecommendAI} in case of a bad outcome, as indicated by the even split between positive and negative satisfaction ratings, while the satisfaction ratings for \textit{ExtendAI} were more differentiated.

Surprisingly, despite the higher effectiveness of \textit{ExtendAI} in improving participants' decision-making, both the respective ratings and interview statements show that participants often perceived \textit{ExtendAI} as less insightful than \textit{RecommendAI}. Apparently, while \textit{ExtendAI} effectively made participants reflect on their decisions, it did so in such a subtle way that participants often did not recognise its contribution to their decision-making, which presents a challenge to the adoption of tools designed like \textit{ExtendAI}.

\subsection{Many Users Want to Think for Themselves}
We designed \textit{ExtendAI} to help users reflect on their reasoning and be cognitively engaged with the task by requiring them to write down their rationale and by giving them feedback without specific ETF recommendations. This expectedly led to higher cognitive effort than with \textit{RecommendAI}, as also confirmed by participants' ratings. Previous studies with crowd workers on simple decision-making tasks found that users prefer AI systems that require less cognitive effort~\cite{bucincaProxyTasksSubjective2020,bucincaTrustThinkCognitive2021,Vasconcelos2023}, pointing towards a trade-off between cognitive engagement and user preference~\cite{bucincaTrustThinkCognitive2021}. The results of our study with a realistic decision-making task are---to a certain extent---contrary to this, as participants were evenly split in their preference for the two AI assistants, often due to differing opinions on the use of specific recommendations. While many participants appreciated the recommendations they received for their actionability, a large portion rejected the recommendations, as they found them to be disengaging and hard to verify. This is in line with other studies on real-world tasks with experienced participants, such as clinicians~\cite{yildirimMultimodalHealthcareAI2024} or pilots~\cite{zhangRecommendationsBackwardForward2024}, who tend to be sceptical toward recommendations. Thus, our findings suggest there are two distinct user groups: one group that welcomes working with ready-to-use AI recommendations, and one group that rejects AI recommendations, as they want to remain as involved in the decision-making as possible.

\subsection{From Trust \& Reliance Calibration to Integration With Users' Decision-Making}
Recent research has focused on trust~\cite{mehrotraSystematicReviewFostering2024} and/or reliance~\cite{schaschekThoseRelyTaxonomy2024} calibration as the core challenge to effective AI-assisted decision-making. While the wide range of work on trust/reliance calibration shows the complexity of the topic, it is still a somewhat simplistic perspective on AI-assisted decision-making that is mostly outcome-focused and often neglects the decision-making process leading up to the outcome. This is partly due to (and further reinforced by) the widespread use of simple input-output mapping tasks in AI-assisted decision-making research~\cite{laiScienceHumanAIDecision2023}. These tasks are often more representative of machine learning tasks than human tasks and afford little room to explore human decision-making processes. By using a more realistic and complex task, our study highlights a range of challenges in AI-assisted decision-making beyond just recognising when the AI is correct or not, related to how AI integrates into and contributes to users' decision-making process. 

Our findings highlight that embedding AI into human decision-making can lead to new tensions, where a benefit of a system is often also a drawback. We observed three such tensions: (1) actionability vs. cognitive engagement, (2) new insights vs. consistency with the user's reasoning, (3) not too early vs. not too late in the decision-making process. We intentionally designed \textit{ExtendAI} to be quite different from \textit{RecommendAI}, for instance by avoiding specific recommendations in \textit{ExtendAI}'s feedback. To an extent, this certainly provoked the tensions we observed, as both assistants tended to land on opposite sides of these spectrums. Yet, while it is conceivable to resolve some of the tensions (e.g., AI advice simultaneously being consistent with users' reasoning and adding novel insights on top), for other tensions, the goals appear to contradict each other (e.g., AI advice should be specific to be actionable, but also not too specific to keep users cognitively engaged), requiring trade-offs appropriate for the application context.

We designed \textit{ExtendAI} to integrate into users' decision-making, and as indicated by both participants' interview statements as well as their ratings (higher \textit{consideration of suggestion}, lower \textit{interference with own decision-making}), \textit{ExtendAI} indeed appears to enable a better integration than \textit{RecommendAI}. Interestingly, participants' remarks suggest that through this integration, \textit{ExtendAI} may `circumvent' or mitigate the issue of trust/reliance calibration to some degree: Since the feedback is embedded into users' own writing and is on a more fine-grained level than concrete recommendations, it is easier to make sense of. As a result, users seem to have less difficulty deciding whether or not to trust the AI, to the extent that the question of trust may not even be users' main concern. This is also reflected by participants' trust ratings, where \textit{ExtendAI} completely lacks negative ratings in contrast to \textit{RecommendAI}, while receiving a large portion of neutral and highly positive ratings.

This benefit of \textit{ExtendAI} came at the cost that users had to write down their rationale first, which was found to be burdensome, although some participants also explicitly recognised the value of this step. Participants made suggestions for how this unwanted cognitive effort could be minimised, for example, by allowing users to formulate their intended trades through direct manipulation, and using natural language only to explain why they want to perform certain trades. However, less obviously, \textit{RecommendAI} introduced cognitive effort as well through the need to evaluate its recommendations, as highlighted by some participants. Unlike \textit{ExtendAI}, this cognitive effort in \textit{RecommendAI} is not primarily the result of interaction design in need of improvement. Instead, it is inherent to the paradigm of recommending complete solutions, and the effort is higher for more complex decisions~\cite{endsleyHereAutonomyLessons2017,simkuteIroniesGenerativeAI2024}. In short, ExtendAI mainly added cognitive effort \textit{before} interacting with the AI but reduced it \textit{afterwards}, whereas for RecommendAI it was the other way around.

Taken together, the findings showed that designing AI decision support systems that build upon people's decision rationales provides an alternative interaction paradigm that could help better integrate AI decision support in human decision-making---even if the instantiations might need to differ and be adapted to the specific task and types of users (e.g. how they provide their rationale to the AI and how detailed and elaborate it should be).

\subsection{Limitations \& Future Work}
Our intention was to use ETF investment as a use case to explore how LLMs can augment human decision-making in general. Nevertheless, certain aspects of our results may have been shaped by the specifics of our task. We argue that the tensions that we identified are likely to be present for most, if not all, complex decision tasks. However, the impact of the advantages and disadvantages of the two types of designs—represented by our two assistants—will likely vary depending on the tasks. For example, in our task, decisions were multifaceted, as participants had to select multiple ETFs and decide on the amount of shares to buy or sell. Consequently, participants could selectively rely on the assistants' suggestions. In contrast, for a task where a single decision has to be made (e.g., medical diagnosis), a recommendation-based approach in particular is more likely to result in a binary ``accept or reject'' situation, which may highlight the disadvantages of this approach more strongly, as is often the case in related work~\cite{zhangRethinkingHumanAICollaboration2024,kawakamiWhyCareWhat2022,blombergActingAnalyticsAccuracy2018}. Similarly, while our participants did discuss the risk of overreliance due to a lack of engagement with \emph{RecommendAI}, it may have been less present than in other studies~~\cite{Bussone2015,jacobsHowMachinelearningRecommendations2021,zhangRecommendationsBackwardForward2024}, since participants had already deeply engaged with the task in the first time step (familiarisation phase) before interacting with our AI assistants. Future work should investigate how analogous AI designs compare in the context of other realistic and complex decision-making tasks.

Another direction for future work is to investigate more quantitative measures with a larger sample size, to explore for example how the perceptions of comparable AI designs depend on personality traits such as need for cognition or the propensity to trust in AI, or factors like users' expertise. It would also be interesting to compare the effect of both types of AI assistance on decision quality in more detail. For our study, we intentionally chose an open-ended task without any objectively correct or false decisions. While this reflects many real-world tasks and allowed us to gain interesting qualitative insights, it also meant that it was difficult to establish clear performance metrics.

Future work could also explore various extensions and modifications to our \emph{ExtendAI} concept, such as allowing multi-turn interactions, or prompting the LLM to be more critical, as some participants felt that \textit{ExtendAI} was overly positive in its tone, reflecting the general behaviour of the underlying GPT-4o model.
Furthermore, different different \textit{forms} of feedback could be evaluated. In our study we limited the feedback to pointing out to the user how they could improve their rationale and drawing their attention to further aspects they could consider in it rather than pointing out specific solutions to them.
Lastly, our study demonstrated the significance of the role that an LLM assumes in the decision-making process. We explored two possible roles with \emph{RecommendAI} and \emph{ExtendAI}, leaving ample room for future work to explore additional roles that could combine elements of both.

\section{Conclusion}
We explored two roles for LLMs in AI-assisted decision-making that leverage their unique capabilities to support complex and multi-faceted decision tasks: (1) a more traditional role where the AI provides the user with direct suggestions, referred to as \emph{RecommendAI}, and (2) an AI that supports users by giving feedback on and extending their own reasoning, referred to as \emph{ExtendAI}. Our findings show that the ExtendAI generally made people feel supported and more reflective in their own thinking and helped them achieve better outcomes. On the other hand, the recommendation-based approach required less effort and seemed to help people get inspiration for new directions. Both of these benefits also came with their respective drawbacks. Therefore, both roles seem to have their pros and cons, leading to several tensions---for example, when there is a need to receive actionable advice while at the same time also being able to think for oneself. These tensions point towards an interesting design space between recommendation-centric approaches and those that build upon people's own reasoning. With this paper, we intend to inform and inspire further exploration of this space.

\begin{acks}
We would like to thank Warren Park for his support with the implementation of the trading platform. We further thank our participants for their valuable input. We also thank our anonymous reviewers for their thoughtful reviews, which helped us to strengthen the paper significantly. The research reported in this work was supported by the EU H2020 ICT48 project ‘Humane AI Net’ under contract \# 952026. The support is gratefully acknowledged.
\end{acks}

\bibliographystyle{ACM-Reference-Format}

\appendix

\section{AI Input and Output Examples}
\label{app:ai-examples}
\subsection{\textit{RecommendAI} Example Output (ER-10)}
\label{app:recommend-examples}
Here is a portfolio recommendation based on your investment profile:
\\
\\
1. VOO (Vanguard S\&P 500 ETF): You already hold 8 shares of this ETF, which is a great choice for broad U.S. market exposure. Given your long-term horizon and cautious approach, maintaining or slightly increasing your investment in VOO can provide stability and growth.
\\
2. VGK (Vanguard FTSE Europe ETF): This ETF offers exposure to European markets, providing geographical diversification. Investing in VGK can help balance your portfolio with international exposure.
\\
3. ICLN (iShares Global Clean Energy ETF): As you are willing to take some risks, adding exposure to the clean energy sector can be a strategic move. ICLN focuses on global clean energy companies, aligning with the growing trend towards renewable energy.
\\
4. ACWV (iShares MSCI Global Min Vol Factor ETF): This ETF aims to provide exposure to global equities with a focus on minimizing volatility. ACWV can add a layer of stability to your portfolio, balancing the riskier elements.
\\
\\
Invest the largest amount of your portfolio into VOO for stability and growth. Allocate a significant portion to VGK for geographical diversification. Consider a moderate investment in ICLN to tap into the clean energy sector's potential. Finally, allocate a smaller portion to ACWV for added stability.

\normalfont

\subsection{\textit{ExtendAI} Example Input/User Rationale (RE-05)}
\label{app:augment-input-examples}
I want to invest 2000\$ into IXN, because of the AI boom, 2000\$ into ICLN because the clean energy boom, and 2000\$ into IXJ because of the the healthcare boom.
\\
\\
I want to invest 2000\$ into EEM and 2180\$ into VOO to diversify across more sectors.  

\normalfont

\subsection{\textit{ExtendAI} Example AI Output/Rationale Extension (RE-05)}
\label{app:augment-output-examples}

\textbf{Here is your edited investment rationale with suggestions for improvement:}
\\
\\
I want to invest \$2000 into IXN, because of the AI boom. \textbf{While IXN provides substantial exposure to the technology sector, it's important to note that it is heavily concentrated in a few large tech companies, which can increase volatility. Consider balancing this with other sectors or regions to mitigate risk.}
\\
\\
I want to invest \$2000 into ICLN because of the clean energy boom. \textbf{ICLN is a great choice for exposure to the clean energy sector, but it is also quite concentrated in a few companies and countries. Diversifying within the clean energy sector or balancing with other sectors could be beneficial.}
\\
\\
I want to invest \$2000 into IXJ because of the healthcare boom. \textbf{IXJ offers good exposure to the healthcare sector, which is typically less volatile and can provide stability. However, it is heavily tilted towards U.S. companies. Consider adding international healthcare ETFs to diversify further.}
\\
\\
I want to invest \$2000 into EEM and \$2180 into VOO to diversify across more sectors. \textbf{EEM offers broad exposure to emerging markets, which can be volatile but offer growth potential. VOO is a solid choice for U.S. large-cap exposure, but it is already part of your holdings. Ensure that your overall portfolio is balanced across different sectors and regions to avoid over-concentration.}
\\
\\
\textbf{Overall, your investment rationale shows a good understanding of current market trends and the desire to diversify. However, consider the concentration risks associated with sector-specific ETFs like IXN, ICLN, and IXJ. Balancing these with more diversified ETFs across different regions and sectors can help mitigate risks. Additionally, since you already hold VOO, ensure that adding more doesn't overly concentrate your portfolio in U.S. large-cap stocks.}
\\
\\
\textbf{These above are just a few things for you to consider for your intended trades. You can now proceed with the trades and the trade button is now enabled. It is up to you if and to what extent you want to consider any of the feedback/considerations provided above.}

\normalfont

\section{AI Assistant Prompts}
\label{app:prompts}

\subsection{\textit{RecommendAI} Prompt}
You are a critical financial advisor embedded into an ETF investing platform. The user is asking for a portfolio recommendation based on their investment profile and goals. 
\\
\\
The ETFs that are available to the user on the investing platform are provided in the attached JSON and the ETFs the user has previously invested in are shown below.
\\
\\
Provide a recommended portfolio of 2-4 ETFs that align with the user's profile, balancing the trade-off between risk and potential return. Ensure that you only recommend ETFs available to the user, as you can see in the JSON below. 
\\
\\
Make sure to reflect and give feedback on the ETFs the user already owns and always suggest a minimum of two ETFs the user does not own yet. Make sure to only give advice for investing in ETFs. Give suggestions for specific ETFs that the user has not mentioned or owns yet as well as advice on selling/buying more of the ETFs the user owns already.
\\
\\
Give a subtle indication of the amount of ETFs that should be sold/bought, but do not give the amount in \$. E.g., `invest the largest amount of your portfolio into ETF XXX...'. Do this after listing the ETFs to invest in.
\\
\\
The profile: `You are a 40-year-old with a stable job and looking to invest in ETFs for a long-term horizon of about 10 years. You are willing to take some risks but you are generally rather cautious. Thus, your goal is to create a diversified ETF profile covering different regions and industries.'
\\
\\
Ensure every suggestion you provide is in bold and the structure is consistent as follows (format all important recommendations using <strong> HTML tags):
\ttfamily
\\
\\
<html>
\\
<body>
\\
<p>Here is a portfolio recommendation based on your investment profile:</p>
\\
(...)
\\
<p>These above are just a few things for you to consider for your intended trades. You can now proceed with the trades and the trade button is now enabled. It is up to you if and to what extent you want to consider any of the feedback/considerations provided above.</p>
\\
</body>
\\
</html>
\\
\\
\normalfont
Make sure the response is consistent with no contradictions.

\subsection{\textit{ExtendAI} Prompt}
You are a critical financial advisor embedded into an ETF investing platform that gives feedback on users' rationales when deciding which ETFs to invest in. Let's assume the user of this system is a beginner investor investing \$20,000--\$30,000 in a portfolio of up to 10 different ETFs that align with their investor profile. The user is looking to invest in ETFs for a long-term horizon of about 10 years they are willing to take some risks but you are generally rather cautious. Thus, the goal is to create a diversified ETF profile covering different regions and industries.
The user has formulated an investment thesis to motivate the selection of ETFs that they plan to invest in. 
\\
\\
The ETFs that are available to the user on the investing platform are provided in the attached JSON and the ETFs the user has previously invested in are shown below.
\\
\\
Make sure that the user listed at least 2 ETFs and has provided a rationale for investing in them. If not, or the user has not input any message, respond with ``Please choose at least 2 ETFs which you would like to invest in.'' or ``You have listed a selection of ETFs that you intend to invest in but you have not yet included a clear motivation/rationale for investing your money in them. Make sure you also include your rationale in the text box below.''
\\
\\
If the user input does list 2 or more ETFs but does not provide any rationale for investing in them, respond with the following: “You have listed a selection of ETFs that you intend to invest in but you have not yet included a clear motivation/rationale for investing your money in them. Make sure you also include your rationale in the text box below.”
\\
\\
When making suggestions on how the user's rationale could be improved, make use of your own knowledge of the returns and volatility of the different ETFs as well as the data in the JSON---in particular, how diversified the ETFs are with regards to ``sector\_holdings'', ``company\_holdings'', ``country\_holdings''. Given the user's goal to adequately balance potential returns and risk, use this trade-off as the main criteria for your suggestions and feedback. Keep the ETFs the user already owns in mind. Make sure to only give advice for investing in ETFs. 
\\
\\
To provide your suggestions, take the user's original rationale and seamlessly ``weave'' your suggestions into their original rationale text (similar to how an editor might work on an author's text). Format your output as HTML and ensure that your feedback and suggestions are highlighted in bold using the <b> HTML tag. 
\\
\\
Here is the format that your response should follow:
\\
1. Start with a brief introduction indicating that this is the edited investment rationale.
\\
2. Provide the original rationale followed by your suggestion in bold within the same paragraph. 
\\
3. End with an overall feedback paragraph summarizing the strengths and concerns of the user’s portfolio along with your suggestions. Formulate it as things to consider for the user.
\\
4. Add the following after the feedback paragraph: ``These above are just a few things for you to consider for your intended trades. You can now make proceed with the trades and the trade button is now enabled. It is up to you if and to what extent you want to consider any of the feedback/considerations provided above.''
\\
\\
Ensure every suggestion you provide is in bold and the structure is consistent as follows:
\ttfamily
\\
\\
<html>
\\
<body>
\\
<p>Here is your edited investment rationale with suggestions for improvement:</p>
\\
<p>Original rationale. <b>Your suggestion.</b></p>
\\
<p>Original rationale. <b>Your suggestion.</b></p>
\\
(...)
\\
<p><b>Overall feedback paragraph summarizing strengths and concerns and providing suggestions.</b></p>
\\
<p>These above are just a few things for you to consider for your intended trades. You can now proceed with the trades and the trade button is now enabled. It is up to you if and to what extent you want to consider any of the feedback/considerations provided above.</p>
\\
</body>
\\
</html>

\normalfont

\onecolumn
\section{Participant Details}
\label{app:participants}
\begin{table}[h]
	\centering
 \caption{Participant sample: Participant RE-2 was recruited through Prolific. Participant RE-1 was not included due to a malfunction of the prototype. CoR stands for Country of Residence. The last two columns refer to the participant's preference and the level of detail their rationales showed.}
 \label{tab:participants}
    \fontsize{8}{5}\selectfont
    \begin{tabular}{lccccccccl}
        \toprule
        \multicolumn{1}{c}{\textbf{Part.ID}} & \textbf{Edu./Degree} & \textbf{Occupation} & \textbf{Gender} & \textbf{Age} & \textbf{CoR} & \textbf{ETF Exper.} & \textbf{AI Preference} & \textbf{Task Time}\\
        \cmidrule{1-9}
            ER-1 & Graduate    & Head of Technology \& AI & Male   & 40 & Germany     & 2 years   & Extend   & 49.55 \\
            ER-2 & Graduate    & Designer                 & Male   & 33 & Germany     & 4 years   & Extend   & 39.83 \\
            ER-3 & Bachelor    & Graduate Student         & Female & 28 & Canada      & 7 years   & Recommend & 49.87 \\
            ER-4 & Graduate    & Technology Lead          & Male   & 32 & Switzerland & 12 years  & Recommend & 31.40 \\
            ER-5 & Graduate    & Software Engineer        & Female & 38 & Germany     & 4 years   & Extend   & 62.72 \\
            ER-6 & Graduate    & Software Engineer        & Female & 29 & Germany     & 1.5 years & Extend   & 48.68 \\
            ER-7 & Graduate    & Software Engineer        & Male   & 29 & Finland     & 1.5 years & Extend   & 63.98 \\
            ER-8 & Bachelor    & Graduate Student                  & Male   & 28 & Switzerland & 0 years   & Recommend & 53.43 \\
            ER-9 & Bachelor    & Graduate Student                  & Male   & 23 & Switzerland & 2 years   & Extend   & 49.87 \\
            ER-10 & Graduate    & PhD Student              & Male   & 30 & UK          & 0 years   & Extend   & 34.78 \\
            RE-2 & Bachelor    & Supply Chain Planner     & Male   & 25 & UK          & 6 years   & Recommend & 17.98 \\
            RE-3 & Graduate    & Research Assistant       & Male   & 32 & Germany     & 0.5 years & Recommend & 54.23 \\
            RE-4 & Graduate    & SAP Consultant           & Female & 33 & Germany     & 1 year   & Recommend & 35.23 \\
            RE-5 & Graduate    & Founder                  & Male   & 32 & Germany     & 7 years   & Extend   & 35.80 \\
            RE-6 & High school & Undergraduate Student          & Male   & 22 & Germany     & 0 years   & Recommend & 28.10 \\
            RE-7 & Graduate    & Researcher               & Male   & 28 & Germany     & 3 years   & Recommend & 40.70 \\
            RE-8 & Graduate    & Designer                 & Female & 31 & Germany     & 4 years   & Recommend & 59.28 \\
            RE-9 & Graduate    & Postdoc                  & Male   & 30 & USA         & 5 years   & Recommend & 40.42 \\
            RE-10 & Graduate    & Research Scientist       & Male   & 32 & Switzerland & 1 year   & Recommend & 65.17 \\
            RE-11 & Graduate    & Assistant Physician      & Male   & 29 & Germany     & 15 years  & Extend   & 38.77 \\
            RE-12 & Graduate    & Associate Professor      & Male   & 34 & UK          & 4 years   & Extend   & 24.17 \\
        \bottomrule
    \end{tabular}
\end{table}

\end{document}